\documentclass{he_symp}
\usepackage{epsf}

\newcommand{\gcc}{{\rm~g\,cm}^{-3}}

\begin{document}

\title{Cooling of Superfluid Neutron Stars}
\author{
D.G. Yakovlev \inst{1}
\and
O.Y. Gnedin \inst{2}
\and
A.D. Kaminker \inst{1}
\and
A.Y. Potekhin \inst{1}
}
\institute{
Ioffe Physical Technical Institute,
         Politekhnicheskaya 26, 194021 St.~Petersburg, Russia
\and
Space Telescope Science Institute,
         3700 San Martin Drive, Baltimore, MD 21218, USA
\\
{\em  yak@astro.ioffe.rssi.ru,
ognedin@stsci.edu,
kam@astro.ioffe.rssi.ru,
palex@astro.ioffe.rssi.ru
}}
\maketitle

% **********************************************************************
\begin{abstract}
Cooling of neutron stars (NSs)
with the cores composed of neutrons,
protons, and electrons is analyzed.
The main cooling regulators
are discussed: opening of direct Urca process
in a NS central kernel;
superfluidity of nucleons in NS interiors;
surface layers of light (accreted) elements;
strong surface magnetic fields.
An emphasis is paid on the cooling
scenario with strong $^1$S$_0$ pairing of protons
and weak $^3$P$_2$ pairing of neutrons in the NS core,
as well as strong $^1$S$_0$ pairing of
neutrons in the NS crust. 
The theory predicts
three types of isolated cooling middle-aged
NSs with distinctly different properties:
low-mass, slowly cooling NSs;
medium-mass, moderately cooling NSs;
massive, rapidly cooling NSs.
The theory is compared with observations
of 
% eight 
isolated NSs --- 
%%%%%%%%%%%%%%%
%(RX J0822--43,
%1E 1207--52, RX J0002+62, PSR 0656+14, Vela, Geminga,
%PSR 1055--52, and RX J1856--3754), 
%%%%%%%%%%%%%%%
pulsars and radio quiet NSs in supernova remnants.
The constraints on physical properties of NSs
which can be inferred from such a comparison are outlined.
\end{abstract}

% **********************************************************************
%                               TEXT BODY
% **********************************************************************
%%%%%%%%%%%%%%%%%%%%%%%%%%%% Sect. 1 %%%%%%%%%%%%%%%%%%%%%%%%%%%%%%%%%%%
\section{Introduction}
\label{sect-intro}
Cooling of neutron stars (NSs) depends on the properties of dense matter
in their crusts and
cores. These properties are still poorly known
and cannot be described unambiguously by contemporary theories.
For instance, calculations
of the equation of state (EOS) of the NS cores
(e.g., Lattimer \& Prakash \cite{lp01})
or the superfluid properties of NS cores and crusts
(e.g., Lombardo \& Schulze \cite{ls01}) give a large
scatter of results which depend on a model of strong interaction
and a many-body theory employed.
However, these properties can be studied by comparing
the results of simulations of NS cooling with the observations
of thermal emission from isolated NSs. We will describe
some recent results of such studies. The history of
NS cooling theory is reviewed, for instance,
by Yakovlev et al.\ (\cite{yls99})

For simplicity, we consider the NS models
with the cores composed mainly of neutrons 
with an admixture of protons and electrons.
We discuss the basic ideas of the cooling
theory, and the main cooling regulators,
paying special attention on the effects
of superfluidity of nucleons in NS interiors.
Our consideration will
be based on recent simulations of the cooling
of superfluid NSs by Kaminker et al.\ (\cite{khy01}),
Potekhin \& Yakovlev (\cite{py01}),
Yakovlev et al.\ (\cite{ykg01}), and Kaminker et al.\
(\cite{kyg02}, hereafter KYG).

% ****    Sect. 2   ******************************************************
\section{Equations of thermal evolution}
\label{therm-balance-transport}
% *************************************************************************

Neutron stars are born very hot in supernova explosions,
with internal temperature $T \sim 10^{11}$ K,
but gradually cool down.
In about 30 s after the birth a star becomes fully
transparent for neutrinos generated
in its interiors. We consider the cooling in the
following neutrino-transparent stage.
The cooling is realized via
two channels, by neutrino emission from the entire stellar body
and by heat transport 
from the internal layers to the surface resulting
in the thermal emission of photons.
For simplicity, we neglect the possible reheating mechanisms
(frictional dissipation of rotational
energy, or Ohmic decay of internal magnetic field,
or the dissipation associated with weak deviations
from the chemical equilibrium as reviewed, e.g., by Page \cite{page98a}).

The internal structure of NSs
can be regarded as temperature-independent
(e.g., Shapiro \& Teukolsky \cite{st83}).
The general relativistic equations of thermal evolution
include the energy and flux equations obtained by
Thorne (\cite{thorne77}). For a spherically symmetric NS,
\begin{equation}
    { 1 \over 4 \pi r^2 {\rm e}^{2 \Phi}} \,
    \sqrt{1 - {2 G m \over c^2 r}} \,
    { \partial  \over \partial  r}
    \left( {\rm e}^{2 \Phi} L_r \right)
    = -Q - {c_v \over {\rm e}^\Phi} \, {\partial T \over \partial t},
\label{therm-therm-balance}
\end{equation}
\begin{equation}
    {L_r \over 4 \pi \kappa r^2} =
    - \sqrt{1 - {2Gm \over c^2 r}} \; {\rm e}^{-\Phi}
     {\partial \over \partial r} \left( T {\rm e}^\Phi \right),
\label{therm-Fourier}
\end{equation}
where $Q$ is the neutrino emissivity
[erg~cm$^{-3}$~s$^{-1}$],
$c_v$ is the heat capacity per cm$^3$ [erg~cm$^{-3}$~K$^{-1}$],
$\kappa$ is the thermal conductivity [erg~cm$^{-1}$~s$^{-1}$~K$^{-1}$],
and $L_r$ is the ``local luminosity'' [erg~s$^{-1}$] defined as the
non-neutrino heat flux transported through a sphere of radius
$r$. The gravitational mass $m(r)$ and the metric function
$\Phi(r)$ are determined by the stellar model.

It is conventional (e.g., Gudmundsson et al.\ \cite{gpe83})
to subdivide the calculation of heat transport
in the neutron-star interior ($r< R_{\rm b}$)
and in the outer heat-blanketing envelope
($R_{\rm b} \le r \le R$), where $R$ is the stellar
radius, and the boundary radius $R_b$
corresponds to the density $\rho_{\rm b} \sim 10^{10}$ g cm$^{-3}$
($\sim$ 100 meters under the surface). 
In the presence of superstrong surface magnetic field, 
it is reasonable to shift $\rho_{\rm b}$ to the neutron drip density,
$4 \times 10^{11}$ g cm$^{-3}$
(Potekhin \& Yakovlev \cite{py01}).
The thermal structure of the blanketing envelope
is studied separately in the stationary,
plane-parallel approximation to relate
the effective surface temperature $T_{\rm s}$ to the temperature
$T_{\rm b}$ at the inner boundary of the envelope.
We discuss the $T_{\rm s}$--$T_{\rm b}$ relation in Sect.\ 3.
It is used as the boundary condition for solving
Eqs.\ (\ref{therm-therm-balance}) and (\ref{therm-Fourier})
at $r<R_{\rm b}$.

The effective temperature determines the photon
luminosity $ L_\gamma = 4 \pi \sigma R^2 T^4_{\rm s}(t)$.
Both, $L_\gamma$ and $T_{\rm s}$ refer to the locally-flat
reference frame on the surface. A distant
observer would register the ``apparent'' luminosity
$L_\gamma^\infty = L_\gamma (1 - r_{\rm g}/R)$
and the ``apparent'' effective temperature
$T_{\rm s}^\infty = T_{\rm s} \, \sqrt{1 - r_{\rm g}/R}$,
where $r_{\rm g}=2GM/c^2$ is the Schwarzschild radius
and $M=m(R)$ is the total gravitational mass.

The main goal of the cooling theory is to calculate
{\it cooling curves}, $T_{\rm s}^\infty(t)$,
to be compared with observations.
One can distinguish three main cooling stages:
(i) the internal relaxation stage ($t \la 10$--100 yr;
Lattimer et al.\ \cite{lattimeretal94}; Gnedin et al.\ \cite{gyp01}),
(ii) the neutrino cooling stage (the neutrino luminosity
$L_\nu \gg L_\gamma$, $t \la 10^5$ yr), and
(iii) the photon cooling stage ($L_\nu \ll L_\gamma$,
$t \ga 10^5$ yr).

After the thermal relaxation, the redshifted
temperature
$ \widetilde{T}(t)= T(r,t)\; {\rm e}^{\Phi(r)}$
becomes constant throughout the stellar interior.
Then Eqs.\ (\ref{therm-therm-balance}) and (\ref{therm-Fourier})
reduce to the equation of global thermal balance,
\begin{eqnarray}
&&   C(\widetilde{T}) \,
     {{\rm d} \widetilde{T} \over {\rm d} t}  =
      - L_\nu^\infty (\widetilde{T}) - L_\gamma^\infty (T_s),
\label{therm-isotherm}\\
&&   L^\infty_\nu (\widetilde{T}) =
     4 \pi \int_0^{R_b}
     { {\rm d}r \, r^2 \, Q (\widetilde{T}) \, {\rm e}^{2 \Phi}
     \over \sqrt{1-2Gm/(c^2r)}},
\label{therm-L_nu} \\
&&   C(\widetilde{T}) =
      4 \pi \int_0^{R_b} {{\rm d}r \, r^2 \, c_v(\widetilde{T}) \over
     \sqrt{1-2Gm/(c^2r)}},
\label{therm-C}
\end{eqnarray}
where $C$ is the total NS heat capacity,
and
$L^\infty_\nu$ is the total neutrino
luminosity.

%%%%%%%%%%%%%%%%%%%%%%%%%%%%% Sect. 3 %%%%%%%%%%%%%%%%%%%%%%%%%%%%%%%
\section{Physics input}
\label{sect-model}
%%%%%%%%%%%%%%%%%%%%%%%%%%%%%%%%%%%%%%%%%%%%%%%%%%%%%%%%%%%%%%%%%%%%%

We describe the results obtained using a fully relativistic
non-isothermal cooling code (Gnedin et al.\ \cite{gyp01}).
The code solves the heat diffusion equations (Sect.\ 2)
in the NS interior.

The physics unput is as follows.
The EOS in the NS crust is taken from Negele \& Vautherin (\cite{nv73})
(assuming spherical atomic nuclei everywhere in the crust).
The core-crust boundary is placed
at the density $
%\rho_{\rm cc}=
1.5 \times 10^{14}$ g cm$^{-3}$.
In the core, we use two phenomenological EOSs
(EOS A and EOS B)
proposed by Prakash et al.\ (\cite{pal88}).

EOS A is model I of Prakash et al.\ (\cite{pal88})
with the compression modulus of saturated
nuclear matter $K=240$ MeV.
EOS B corresponds to $K=180$ MeV
and to the simplified form of the symmetry energy
suggested by Page \& Applegate (\cite{pa92}).
It has been used in many
papers (e.g., Page \& Applegate \cite{pa92}, Yakovlev
et al.\ \cite{yls99,ykgh01}, and references therein).

The masses, central densities, and radii
of two stellar configurations for EOSs A and B are given in Table 1.
The first configuration corresponds to the maximum-mass
NS. The values of $M_{\rm max}$ reveal that EOS A
is stiff, and EOS B is moderate.
The second configuration
has a central density $\rho_{\rm c}$ at which the
direct Urca process switches on
(the process is allowed
at $\rho_{\rm c} > \rho_{\rm D}$,
which corresponds to $M> M_{\rm D}$).
EOS B is based on a smaller symmetry
energy and opens the direct Urca process
at a higher $\rho$.

%%%%%%%%%%%%%%%%%%%%%%%%%%%%%% Table 1 %%%%%%%%%%%%%%%%%%%%%%%%%%%%%%%%%%%%%%%%%%
\begin{table}[t]
\caption{NS models employing EOSs A and B (from KYG)}
\begin{tabular}{|l|l|l|l|}
\hline
Model  & Main parameters                       &     EOS A     &     EOS B     \\
\hline \hline
Maximum& $M_{\rm max}/{\rm M}_\odot$           &  1.977        &  1.73        \\
mass   & $\rho_{\rm cmax}/10^{14}$ g cm$^{-3}$ &  25.75        &  32.5         \\
model  & $R$ km                                &  10.754       &  9.71         \\
\hline \hline
Direct Urca& $M_{\rm D}/{\rm M}_\odot$             &   1.358       &  1.44     \\
threshold& $\rho_{\rm D}/10^{14}$ g cm$^{-3}$ &   7.851       &  12.98        \\
model              & $R$ km                                & 12.98& 11.54      \\
\hline
\end{tabular}
\end{table}
%%%%%%%%%%%%%%%%%%%%%%%%%%%%%%%%%%%%%%%%%%%%%%%%%%%%%%%%%%%%%%%%%%%%%%%%%%%%%%%%

The cooling code includes all important neutrino emission
processes in the NS core (direct and modified Urca processes,
neutrino bremsstrahlung in nucleon-nucleon scattering,
neutrino emission due to Cooper pairing of nucleons) and
in the crust (plasmon decay, neutrino bremsstrahlung due to
scattering of electrons off atomic nuclei, electron-positron
annihilation into neutrino pairs, neutrino emission due
to Cooper pairing of neutrons in the inner crust).
These processes are reviewed, for instance,
by Yakovlev et al.\ (\cite{ykgh01}).
The effective masses of protons and neutrons in the core and
free neutrons in the crust are taken to be 0.7 of the bare
nucleon masses. The values of the thermal conductivity
in the NS crust and core are the same as used by Gnedin
et al.\ (\cite{gyp01}).

The $T_{\rm s}$--$T_{\rm b}$ 
relationship is taken from Potekhin et al.\ (\cite{pcy97})
and Potekhin \& Yakovlev (\cite{py01}).
It is relevant
either for the surface layers made of iron
(without any magnetic field and with
the dipole surface magnetic fields $B \la 10^{15}$ G),
or for the non-magnetic surface layers
containing light elements. 
In addition, we have modified the relationship
by incorporating our new results which take into account
the combined effects of accreting envelopes and magnetic fields.
It is assumed that the
surface magnetic field induces an anisotropic heat transport
in the heat-blanketing envelope but does not violate
the isotropic (radial) heat diffusion in the deeper NS layers.
This preserves the scheme of
solving the radial heat diffusion equations (Sect.\ 2).
In this case, $T_{\rm s}$ varies
over the NS surface, and the cooling theory deals
with the mean surface temperature $\bar{T}_{\rm s}$
that determines the NS photon luminosity.
For brevity, we denote $\bar{T}_{\rm s}$ as $T_{\rm s}$.
It is also assumed that a NS may have a hydrogen atmosphere
even if the heat-blanketing envelope is mostly made of iron.
The majority of cooling curves are calculated for
the non-magnetized heat-blanketing envelopes
made of iron. The exceptions
are discussed in Sect.\ 6.3.

%%%%%%%%%%%%%%%%%%%%%%%%%%%%%% Table 2 %%%%%%%%%%%%%%%%%%%%%%%%%%%%%%%%%
\begin{table}[t]
\caption{Parameters of superfluid models in Eq.\ (\ref{Tc}); from KYG}
\begin{tabular}{lllllll}
%\hline \\
Pair-     & Mo-   & $T_{0}/10^9~$K & $k_0$ &  $k_1$   & $k_2$    &    $k_3$ \\
ing       & del   &          & fm$^{-1}$  & fm$^{-1}$ & fm$^{-1}$& fm$^{-1}$ \\
\hline
$^1$S$_0$ & 1p   & 20.29     &       0    &  1.117    & 1.241    &  0.1473  \\
$^1$S$_0$ & 2p   & 17        &       0    &  1.117    & 1.329    &  0.1179  \\
$^1$S$_0$ & 3p   & 14.5      &       0    &  1.117    & 1.518    &  0.1179  \\
$^1$S$_0$ & 1ns  & 10.2      &       0    &  0.6      & 1.45     &  0.1     \\
$^1$S$_0$ & 2ns  & 7.9       &       0    &  0.3      & 1.45     &  0.01 \\
$^1$S$_0$ & 3ns  & 1800      &       0    &   21      & 1.45     &  0.4125 \\
$^3$P$_2$ & 1nt  & 6.461     &       1    &  1.961    & 2.755    &  1.3
%\\
%\hline
\end{tabular}
\end{table}
%%%%%%%%%%%%%%%%%%%%%%%%%%%%%%%%%%%%%%%%%%%%%%%%%%%%%%%%%%%%%%%%%%%%%%%%%

We will focus on the effects of nucleon superfluidity
on NS cooling. The superfluid properties of NS matter
are characterized by the density-dependent critical temperatures
$T_{\rm c}(\rho)$ of nucleons. Microscopic theories predict
(e.g., Lombardo \& Schulze \cite{ls01}, Yakovlev et al.\ \cite{yls99})
superfluidities of three types: singlet-state ($^1$S$_0$)
pairing of neutrons ($T_{\rm c}=T_{\rm cns}$) in the
inner crust and outermost core; $^1$S$_0$ proton pairing
($T_{\rm c}=T_{\rm cp}$) in the core; and triplet-state
($^3$P$_2$) neutron pairing ($T_{\rm c}=T_{\rm cnt}$) in the core.
Superfluidity of nucleons suppresses neutrino processes
involving nucleons, initiates a specific mechanism
of neutrino emission associated with Cooper pairing
of nucleons (Flowers et al.\ \cite{frs76}),
and affects the nucleon heat capacity. All these
effects are incorporated into the cooling code
as described by Yakovlev et al.\ (\cite{yls99,ykgh01})
and Gnedin et al.\ (\cite{gyp01}).

%%%%%%%%%%%%%%%%%%%%%%%%%%%%%%%%%%%%%%%%%%%%%%%%%%%%%%%%%%%%%%
\begin{figure}
\centering
\epsfxsize=86mm
\epsffile[20 150 570 700]{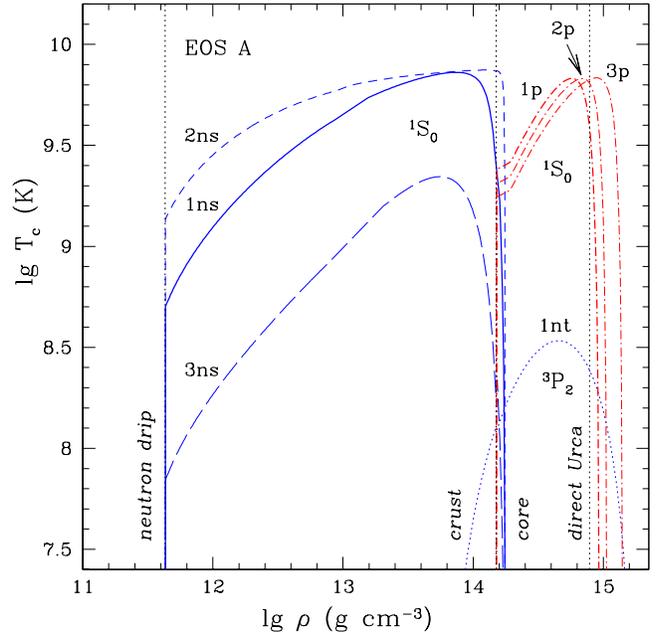}
\caption{
Density dependence (from KYG) of the critical temperatures
for three models 1p, 2p, and 3p
of the proton superfluidity (dots-and-dashes) in the core
(with EOS A); three models
1ns, 2ns, and 3ns of $^1$S$_0$ neutron superfluidity
(solid, short-dashed, and long-dashed lines);
and one model 1nt of $^3$P$_2$ neutron superfluidity (dots)
used in cooling
simulations.
Vertical dotted lines indicate the neutron drip point,
core-crust interface, and direct Urca threshold.
}
\label{fig1}
\end{figure}
%%%%%%%%%%%%%%%%%%%%%%%%%%%%%%%%%%%%%%%%%%%%%%%%%%%%%%%%%%%%%%%

%%%%%%%%%%%%%%%%%%%%%%%%%%%%%%%%%%%%%%%%%%%%%%%%%%%%%%%%%%%%%%
\begin{figure}
\centering
\epsfxsize=86mm
\epsffile[20 35 360 375]{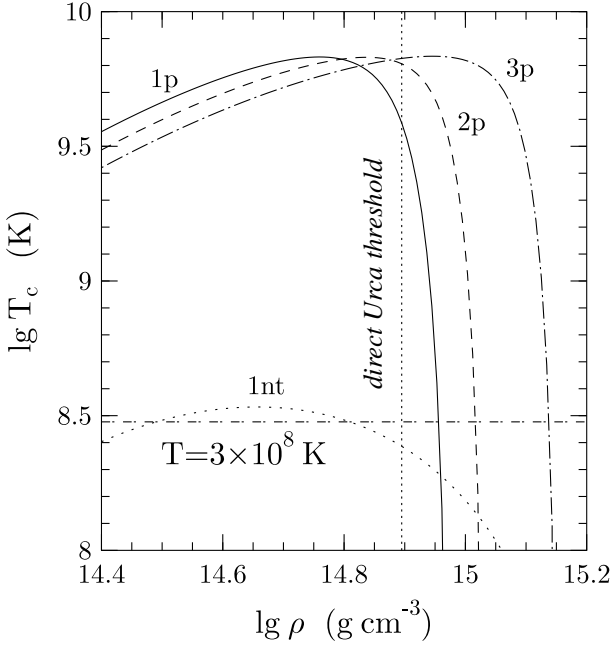}
\caption{
Density dependence of the critical temperatures
of superfluidity of protons (models 1p, 2p, and 3p ---
solid, dashed, and dot-and-dashed lines) 
and neutrons (model 1nt --- dotted line)
in the NS core for EOS A on a larger scale than in  
Fig.\ \ref{fig1}.  Vertical
dotted line indicates direct Urca threshold.
Horizontal line is the temperature of matter,
$T=3 \times 10^8$ K, adopted in Fig.\ \ref{fig3}.
}
\label{fig2}
\end{figure}
%%%%%%%%%%%%%%%%%%%%%%%%%%%%%%%%%%%%%%%%%%%%%%%%%%%%%%%%%%%%%%%

%%%%%%%%%%%%%%%%%%%%%%%%%%%%%%%%%%%%%%%%%%%%%%%%%%%%%%%%%%%%%%
\begin{figure}
\centering
\epsfxsize=86mm
\epsffile[20 35 360 375]{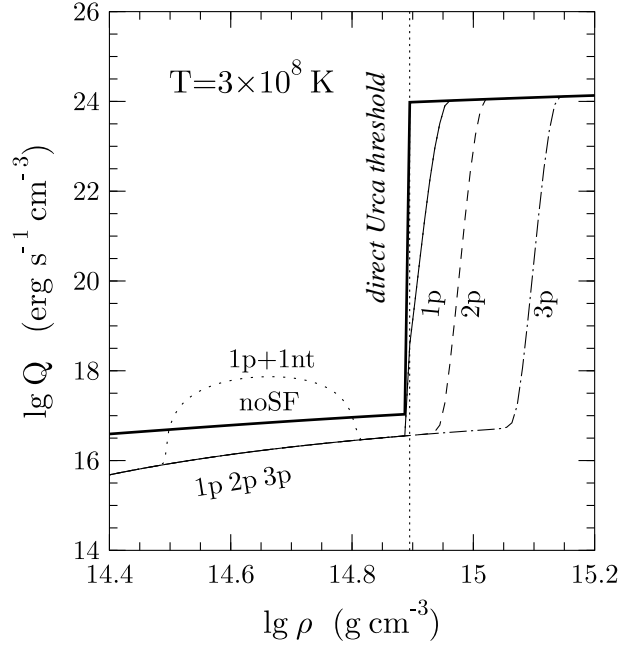}
\caption{
Density dependence of the neutrino emissivity $Q$
in the NS core at $T=3 \times 10^8$ K and EOS A. Solid,
dashed and dot-and-dashed 
lines 1p, 2p, and 3p are plotted for proton superfluidity
models displayed in Fig.\ 2. The dotted line
shows $Q$ in the presence of proton superfluidity 1p
and neutron superfluidity 1nt. The thick solid line
is $Q$ in non-superfluid matter.
}
\label{fig3}
\end{figure}
%%%%%%%%%%%%%%%%%%%%%%%%%%%%%%%%%%%%%%%%%%%%%%%%%%%%%%%%%%%%%%%

%%%%%%%%%%%%%%%%%%%%%%%%%%%%%%%%%%%%%%%%%%%%%%%%%%%%%%%%%%%%%%
\begin{figure}
\centering
\epsfxsize=86mm
\epsffile[20 35 360 375]{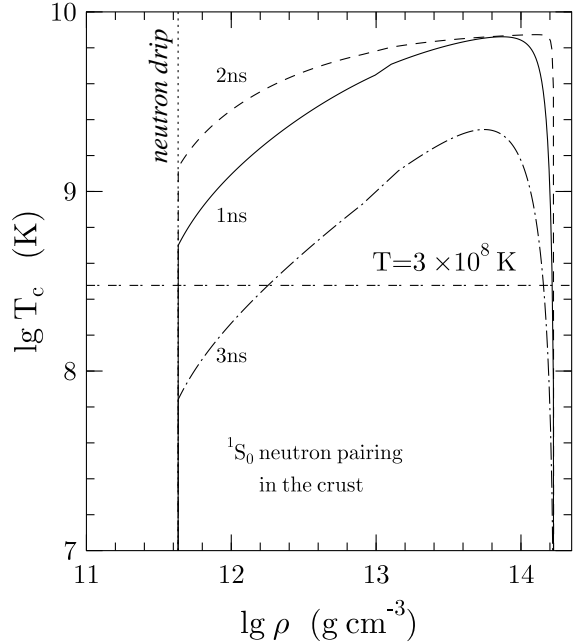}
\caption{
Density dependence of the critical temperatures
of superfluidity of neutrons (models 1ns, 2ns, and 3ns)
in the NS crust (cf.\ Fig.\ 1). Vertical
dotted line indicates neutron drip point.
Horizontal line is the temperature of matter,
$T=3 \times 10^8$ K, adopted in Fig.\ 5.
}
\label{fig4}
\end{figure}
%%%%%%%%%%%%%%%%%%%%%%%%%%%%%%%%%%%%%%%%%%%%%%%%%%%%%%%%%%%%%%%

%%%%%%%%%%%%%%%%%%%%%%%%%%%%%%%%%%%%%%%%%%%%%%%%%%%%%%%%%%%%%%
\begin{figure}
\centering
\epsfxsize=86mm
\epsffile[20 35 360 375]{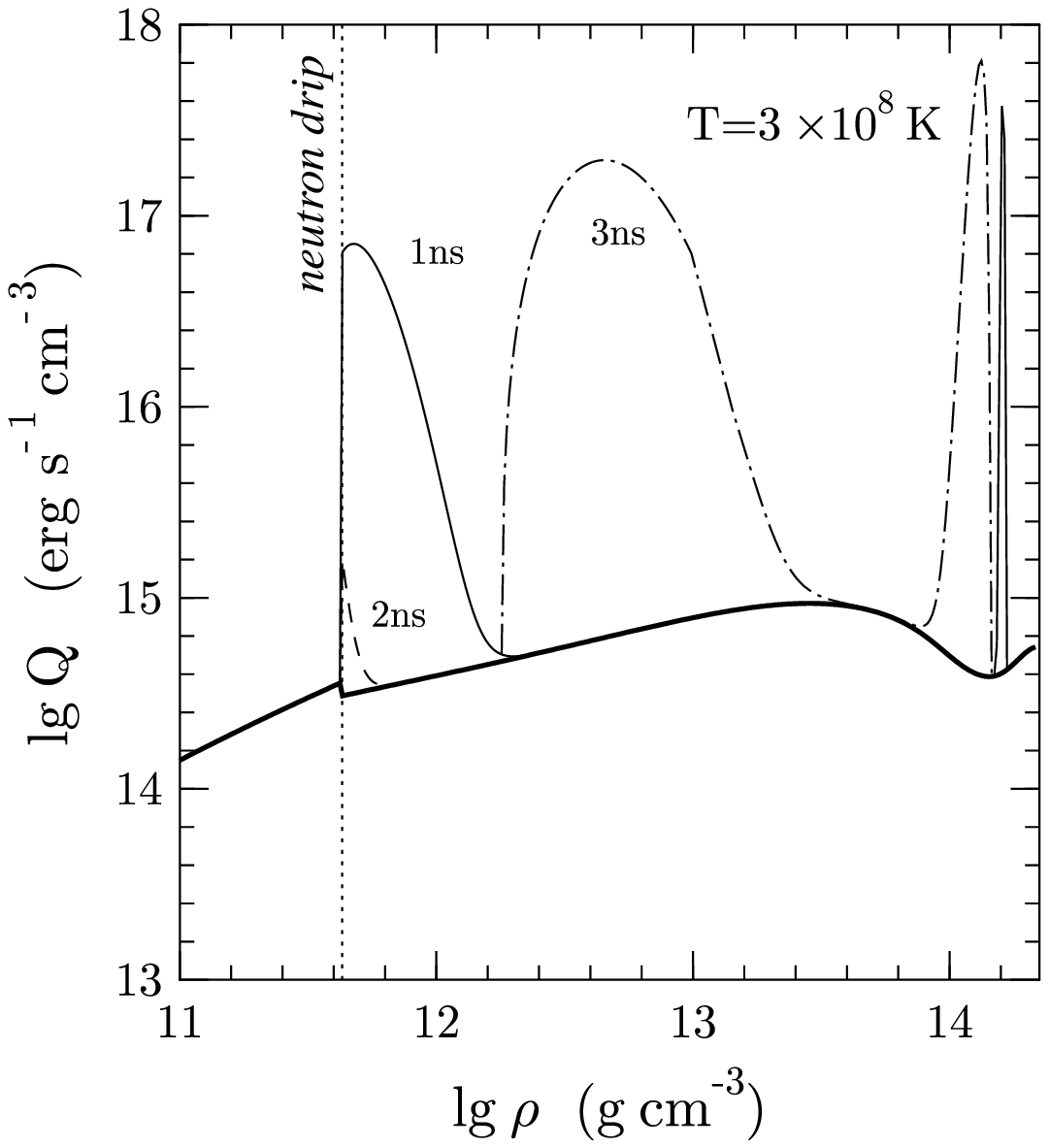}
\caption{
Density dependence of the neutrino emissivity $Q$
in the NS crust at $T=3 \times 10^8$ K. Thin solid,
dashed, and dot-and-dashed
lines 1ns, 2ns, and 3ns are calculated for neutron superfluidity
models displayed in Fig.\ 4.
The thick solid line
is for non-superfluid matter.
The core-crust boundary is artificially shifted to
higher $\rho$ for a better visualization of
the effects of neutrino emission associated with
$^1$S$_0$ Cooper pairing of neutrons.
}
\label{fig5}
\end{figure}
%%%%%%%%%%%%%%%%%%%%%%%%%%%%%%%%%%%%%%%%%%%%%%%%%%%%%%%%%%%%%%%

The microscopic calculations of superfluid critical temperatures
$T_{\rm c}(\rho)$ give a large scatter of results
(e.g., Lombardo \& Schulze \cite{ls01}), but
some common features are clear. For instance, $T_{\rm c}(\rho)$
increases with $\rho$ at sufficiently low densities  % $\rho$
(due to an increasing strength of the attractive
part of nucleon-nucleon interaction),
reaches maximum and then decreases (due to
a short-range nucleon-nucleon repulsion),
vanishing at a rather high density. For $T_{\rm cns}(\rho)$,
the maximum occurs at sub-nuclear densities, while the switch off
takes place at $\rho \sim \rho_0$, where $\rho_0 \approx 2.8 \times 10^{14}$
g cm$^{-3}$ is the saturated nuclear matter density.
For $T_{\rm cp}(\rho)$ and $T_{\rm cnt}(\rho)$,
the maxima take place at a few $\rho_0$ and the fall occurs at
the densities several times higher. The maximum values of $T_{\rm c}$
range from about $10^8$ K (or even lower) to $(2 - 3) \times 10^{10}$ K,
depending on the microscopic theoretical model.
The maximum values of $T_{\rm cnt}$ are typically lower than
those of $T_{\rm cp}$ and $T_{\rm cns}$, due to the weaker nucleon-nucleon
attraction in the $^3$P$_2$ state.

Instead of studying $T_{\rm c}$ as a function of $\rho$, it is often
convenient to consider $T_{\rm c}$ as a function of
the nucleon Fermi wavenumber
$k=k_{\rm F\!N}=(3 \pi^2 n_{\rm N})^{1/3}$, where $n_{\rm N}$
is the number density of nucleon species N=n or p (neutrons or protons).
Moreover, instead of $T_{\rm c}(k)$ one often
considers $\Delta(k)$, the zero-temperature superfluid
gap. For the $^1$S$_0$ pairing, assuming BCS theory,
one has $\Delta(k)=T_{\rm c}(k)/0.5669$.
In the case of $^3$P$_2$ neutron pairing,
the gap depends on the orientation of nucleon momenta with
respect to the quantization axis.
Following the majority of papers, we adopt the $^3$P$_2$
pairing with zero projection of
the total angular momentum on the quantization
axis. In that case $\Delta_{\rm nt}(k)=T_{\rm cnt}(k)/0.8416$,
where $\Delta_{\rm nt}(k)$ is the minimum value of the gap
on the neutron Fermi surface % (realized on
(e.g., Yakovlev et al.\ \cite{yls99}).
The dependence of the gaps on $k$ for the selected
superfluidity models is plotted in KYG.

Taking into account large scatter of theoretical values of
$T_{\rm c}(k)$ it is instructive to consider
$T_{\rm c}(k)$ as unknown functions and analyze (Sect.\ 6)
which of them are consistent with observations.
In this paper, following KYG,
we adopt the parameterization of $T_{\rm c}$ of
the form:
\begin{equation}
    T_{\rm c}= T_0 \, {(k-k_0)^2 \over (k-k_0)^2 + k_1^2}
     \; {(k-k_2)^2 \over  (k-k_2)^2 + k_3^2 }~,
\label{Tc}
\end{equation}
for $k_0< k < k_2$; and $T_{\rm c}=0$, for $k \leq k_0$ or $k \geq k_2$.
The factor $T_0$ regulates the amplitude of $T_{\rm c}$,
$k_0$ and $k_2$ determine
positions of the low- and high-density cutoffs,
while $k_1$ and $k_3$ specify the shape of $T_{\rm c}(\rho)$.
All wave numbers, $k$, $k_0$, \ldots $k_3$ are expressed
in fm$^{-1}$.
KYG verified that by tuning $T_0$, $k_0$, \ldots  $k_3$,
the parameterization accurately describes
numerous results of microscopic calculations.

Following KYG
we use three models of $^1$S$_0$ proton superfluidity,
three models of $^1$S$_0$ neutron superfluidity,
and one model of $^3$P$_2$ neutron superfluidity.
The parameters of the models are given in Table 2,
and the appropriate $T_{\rm c}(\rho)$ are plotted in Figs.\ \ref{fig1},
\ref{fig2}, and \ref{fig4}.
One has $k_0=0$ for $^1$S$_0$ pairing. At any given $\rho$
it is reasonable to choose the neutron
superfluidity ($^1$S$_0$ or $^3$P$_2$)
with higher $T_{\rm c}$.

Models 1ns and 2ns of neutron pairing in the crust
correspond to about the same, rather strong
superfluidity (with maximum $T_{\rm cns} \approx 7 \times 10^9$ K).
Model 2ns has flatter maximum
and sharper decreasing slopes in the wings
(near the crust-core interface and the neutron drip point). Model 3ns
represents a much weaker superfluidity, with maximum
$T_{\rm cns}\approx 2.4 \times 10^9$ K and a narrower
density profile.

The proton superfluidity curves 1p, 2p, and 3p in Figs.\ \ref{fig1} and
\ref{fig2}  are similar.
The maximum values of $T_{\rm cp}$ are about $7 \times 10^9$ K for all
three models. 
% Note that model 1p was used in Papers I and II.
The models differ by the positions of the maximum and decreasing
slopes of $T_{\rm cp}(\rho)$.
The decreasing slope of model 1p
is slightly above the threshold density
of the direct Urca process (for EOS A), while the slopes for models
2p and 3p are shifted to higher $\rho$.
Our models 1p, 2p, and 3p are typical for those
microscopic theories which adopt a moderately
strong medium polarization of
proton-proton interaction.

Finally, the dotted curve in Figs.\ \ref{fig1} and \ref{fig2}
shows $T_{\rm cnt}(\rho)$
for 1nt model of $^3$P$_2$ neutron pairing
(used by Kaminker et al.\ \cite{khy01}). The curve falls within
a wide scatter of $T_{\rm cnt}(\rho)$,
provided by microscopic theories.  % scatter.

%%%%%%%%%%%%%%%%%%%%%%%%%%%%%%  Table 3 %%%%%%%%%%%%%%%%%%%%%%%%%%%%
\newcommand{\rrr}{\rule{0cm}{0.4cm}}
\newcommand{\hh}{\rule{0.5cm}{0cm}}
\newcommand{\hb}{\rule{0.4cm}{0cm}}

\begin{table*}[!t]   % "*" ignores the twocolumn-format if adopted
\caption[]{Surface temperatures of eight isolated middle-aged neutron stars
inferred from observations}
\label{tab-cool-data}
\begin{center}
\begin{tabular}{|| l | l | l | l | c | l ||}
\hline
\hline
% Column titles ==============================================
 Source & lg~t & lg~$T_{\rm s}^\infty$ & Model$^{a)}$ & Confid. & References   \\
              & [yr] &  [K]             &              & level   &           \\
%=============================================================
\hline
\hline
RX$\,$J0822--43 & 3.57 & $ ~~~6.23^{+0.02 \rrr}_{-0.02} $ & H  &  95.5\% &
Zavlin et al.\ (\cite{ztp99}) \\
%                                  Zavlin, Tr\"{u}mper, Pavlov 1999
%\hline
1E$\,$1207--52  & 3.85 & $ ~~~6.10^{+0.05 \rrr}_{-0.06} $  & H  & 90\% &
Zavlin et al.\ (\cite{zpt98}) \\
%                                  Zavlin, Pavlov, Tr\"{u}mper 1998
%\hline
RX$\,$J0002+62 & $3.95^{b)}$ & $ ~~~6.03^{+0.03 \rrr}_{-0.03} $ & H & 95.5\%&
Zavlin \& Pavlov (\cite{zp99}) \\
%                                  Zavlin, Pavlov 1999
%\hline
PSR~0833--45 (Vela) & $4.4^{c)}$ & $ ~~~5.83^{+0.02 \rrr}_{-0.02}$ & H & 68\%  &
Pavlov et al.\ (\cite{pavlovetal01})\\
%      Pavlov, Zavlin, Sanwal, Burwitz, Garmire 2001
%\hline
PSR~0656+14 & 5.00 & $ ~~~5.96^{+0.02\rrr}_{-0.03} $ & bb &  90\%  &
Possenti et al.\ (\cite{pmc96}) \\
%              Possenti, Mereghetti, Colpi 1996
%\hline
PSR~0633+1748 (Geminga) & 5.53 & $ ~~~5.75^{+0.05\rrr}_{-0.08} $ & bb & 90\% &
Halpern \& Wang (\cite{hw97}) \\
%                                Halpern and Wang 1997
%\hline
PSR~1055--52 & 5.73 & $ ~~~5.88^{+0.03\rrr}_{-0.04} $ & bb & $^{d)}$ &
\"{O}gelman (\cite{ogelman95}) \\
%                                Ogelman 1995
%\hline
RX~J1856--3754 & 5.95 & $ ~~~5.72^{+0.05\rrr}_{-0.06} $
& $^{e)}$  & $^{d)}$ &
Pons et al.\ (\cite{ponsetal02}) \\
%                                Pons et al 2002
\hline
\end{tabular}
\begin{tabular}{l}
  $^{a)}\,$\rrr{\footnotesize Observations are interpreted either with
a hydrogen atmosphere model (H), or with a black body spectrum (bb)}\\[0.5ex]
  $^{b)}\,$\rrr{\footnotesize The mean age taken according
 to Craig et al.\ (\cite{chp97}).}\\[-0.5ex]
%                   Craig, Hailey, Pisarski 1997
  $^{c )}\,$\rrr{\footnotesize According to
        Lyne et al.\ (\cite{lyneetal96}).}\\[-0.5ex]
%                   Lyne, Pritchard, Graham-Smith, Camilo 1996
  $^{d)}\,$\rrr{\footnotesize Confidence level is uncertain.}\\[0.5ex]
  $^{e)}\,$\rrr{\footnotesize Analytic fit with Si-ash atmosphere model
     of Pons et al.\ (\cite{ponsetal02}).} %\\[0.5ex]
%
%  $^{g)}\,$\rrr{\footnotesize Estimated error for a joint fit of the optical
%               plus X-ray data.}\\[0.5ex]
\end{tabular}
\end{center}
\end{table*}

% Section 4 %%%%%%%%%%%%%%%%%%%%%%%%%%%%%%%%%%%%%%%%%%%%%%%%%%%%%
\section{Main regulators of neutron-star cooling}
%%%%%%%%%%%%%%%%%%%%%%%%%%%%%%%%%%%%%%%%%%%%%%%%%%%%%%%%%%%%%%%%%

Theoretical cooling curves depend on many features of
NS models. Below we discuss four main regulators
of the cooling of middle-aged NSs: 
(1) central kernels, where neutrino emission
is enhanced by direct Urca process;
(2) the effects of nucleon superfluidity
on neutrino emission; 
(3) surface envelopes of light elements;
and (4) surface magnetic fields.
The effects of these factors are illustrated
in Figs.\ \ref{fig2}--\ref{fig7}.
The effects (1) and (2) are the strongest,
although (3) and (4) can also be important.

%%%%%% Section 4.1 %%%%%%%%%%%%%%%%%%%%%%%%%%%%%%%%%%%%%%%%%
\subsection{Direct Urca process in non-superfluid NSs}
%%%%%%%%%%%%%%%%%%%%%%%%%%%%%%%%%%%%%%%%%%%%%%%%%%%%%%%

The NS cooling is strongly affected by
the presence of the central NS kernels ($\rho > \rho_{\rm D}$,
Table 1), where direct Urca process --
the most powerful neutrino emission mechanism -- is open.
The effect is especially pronounced in the absence
of nucleon superfluidity. For instance, Fig.\ \ref{fig3}
shows the density profile of the total neutrino emissivity
throughout the NS core at a temperature $T= 3 \times 10^8$ K.
The emissivity of non-superfluid matter is shown by
the thick line. The jump by about 7 orders of magnitude
is associated with the direct Urca threshold, $\rho_{\rm D}$.
The NSs with the central density $\rho_{\rm c} < \rho_{\rm D}$
have no kernels with the enhanced neutrino emission and
show {\it slow} cooling. The NSs with $\rho_{\rm c} > \rho_{\rm D}$
will have these kernels and show {\it fast} cooling.
The presence of the kernels
has dramatic effects on the NS cooling
(Lattimer et al.\ \cite{lpph91}).
%%%%%%%%%%%%%%%%%%%%%%%%%%%%%%%%%%%%%%%%%%%%%%%%%%%%%%%%%%%%%%%%%%%
 
%%%%%%% Sect. 4.2 %%%%%%%%%%%%%%%%%%%%%%%%%%%%%%%%%%%%%%%%%%%%%%%%%%%%
\subsection{The effects of superfluidity on neutrino emission}
%%%%%%%%%%%%%%%%%%%%%%%%%%%%%%%%%%%%%%%%%%%%%%%%%%%%%%%%%%%%%%%%%%%%%%

First we discuss the effects of superfluidity in the NS
cores. The adopted models of proton and neutron critical
temperatures (Sect.\ 3) in the core are shown in 
Figs.\ \ref{fig1} and \ref{fig2}.
%%%%%%%%%%%%%%
%on a larger scale than in Fig.\ 1. 
%%%%%%%%%%%%%%
The appropriate
neutrino emissivities are displayed in Fig.\ 3.
A proton superfluidity 
%%%%%%%%%%%%
%(lines 1p, 2p, or 3p,
%in Figs.\ 2 and 3) 
%%%%%%%%%%%%%
drastically changes
the density profiles of the neutrino emissivity, $Q(\rho)$,
as compared with the non-superfluid case
(lines 1p, 2p, and 3p in Figs.\ \ref{fig2} and \ref{fig3}).

At not too high density, $\rho \la \rho_{\rm I}$
(the threshold density $\rho_{\rm I}$ is specified in Sect.\ 6),
a strong proton superfluidity
almost switches off all neutrino processes
involving protons, particularly, modified Urca process and
even the most powerful direct Urca process.
In this regime, the main neutrino
emission is provided by neutrino-pair bremsstrahlung
in neutron-neutron scattering;
its level is {\it lower} than in a slowly cooling non-superfluid
NS regulated by non-suppressed modified Urca process.
Since strong proton superfluidity 3p extends to higher $\rho$,
this regime persists to higher 
$\rho$, for this superfluidity.

At rather high $\rho \ga \rho_{\rm II}$ 
(the value $\rho_{\rm II}$ is also specified in Sect.\ 6)
the proton superfluidity
dies out and we have the neutrino emissivity fully
enhanced by direct Urca process. Again, $\rho_{\rm II}$ is
higher for proton superfluidity 3p.

At $\rho_{\rm I} \la \rho \la \rho_{\rm II}$
the neutrino emissivity is determined by direct Urca
process partly suppressed by proton superfluidity.
Evidently, this regime is absent in non-superfluid NSs.
In this regime, the emissivity is sensitive to
the decreasing slope of the $T_{\rm cp}(\rho)$ profile and
to EOS in the NS core. 
% The limiting densities $\rho_{\rm I}$
% and $\rho_{\rm II}$ are also sensitive to these factors,
% and they depend (not too strongly) on the internal stellar temperature
% (i.e., on NS age). 
%%%%%%%%%%%%%%%%%
%These densities will be discussed
%in Sect.\ 6 in more detail.
%%%%%%%%%%%%%%%%%

Thus, in any $Q(\rho)$ profile, we can distinguish three
distinctly different parts: (I) $\rho \la \rho_{\rm I}$;
(II) $\rho_{\rm I} \la \rho \la \rho_{\rm II}$;
(III) $\rho \ga \rho_{\rm II}$. 
They will lead (Sect.\ 6) to three types of cooling superfluid NSs.

Now we switch on neutron superfluidity in the core
(model 1nt) in addition to the proton one.
For $T=3 \times 10^8$ K, this superfluidity
occurs in a density range from $\sim 3 \times 10^{14}$
g cm$^{-3}$ to $\sim 6 \times 10^{14}$ g cm$^{-3}$ (Fig.\ 2).
It creates a powerful splash of neutrino emission associated
with Cooper pairing of neutrons as shown in Fig.\ 3 by
the dotted line. This splash is actually the same
for any model of proton superfluidity (1p, 2p, or 3p).

In addition, we discuss the effects of the neutron
superfluidity in the crust. The density profiles of
the critical temperature $T_{\rm cns}(\rho)$
(models 1ns, 2ns, and 3ns; Table 2; Figs.\ \ref{fig1} and \ref{fig4}) 
and the appropriate neutrino emissivities $Q(\rho)$ at
$T=3 \times 10^8$ K are shown in Fig.\ \ref{fig5}.
The thick line in Fig.\ \ref{fig5} exhibits the emissivity
of the non-superfluid crust provided mainly by
neutrino-pair bremsstrahlung in electron-nucleus
scattering. One can observe strong additional peaks
of neutrino emission from superfluid crust
produced by Cooper pairing
of neutrons. The peaks are very sensitive to the
models of $T_{\rm cns}(\rho)$.
However, since the NS crust contains
a small fraction of NS mass (about 1--5 \% or smaller,
for the adopted NS models)
this additional neutrino emission is important only
in those NSs (Sect.\ 6.3) whose cores produce low neutrino luminosity.
% We will discuss these cases in Sect.\ 6.3 in more details.

%%%%%%% 4.3 %%%%%%%%%%%%%%%%%%%%%%%%%%%%%%%%%%%%%%%%%%%%%%%%%%%%%
\subsection{Accreted surface layers}
%%%%%%%%%%%%%%%%%%%%%%%%%%%%%%%%%%%%%%%%%%%%%%%%%%%%%%%%%%%%%%%%%

A heat-blanketing envelope (Sect.\ 2) is traditionally assumed
to be made of iron (e.g., Gudmundsson et al.\ \cite{gpe83}).
However, the envelope may partly
contain light elements (H, He) provided, for instance,
by accretion. 
In the heat-blanketing layers, the
heat is mainly carried by degenerate electrons.
The thermal conductivity of
electrons which scatter off light ions is higher
than the conductivity limited by scattering off iron
ions. Thus, the accreted envelope is more
heat transparent. The appropriate $T_{\rm s}-T_{\rm b}$ relationship
was calculated by Potekhin et al.\ (\cite{pcy97}).
It depends on $\Delta M$, the mass of accreted
hydrogen or helium. The calculation takes into account
nuclear burning of light elements in a hot and/or dense
plasma, i.e., $\Delta M$ is the total accreted mass
which may partly be processed into heavier elements
due to nuclear transformations in accreted matter. The dependence of
the photon NS luminosity on $\Delta M$ for four values
of the temperature 
at the bottom of the heat-blanketing envelope 
($T_{\rm b}=3\times 10^7$, 
$10^8$, $3\times 10^8$, and $10^9$ K) is shown
in Fig.\ 6. If the heat-blanketing envelope is fully
replaced by accreted material, the NS luminosity can
increase by about one order of magnitude, i.e.,
the surface temperature $T_{\rm s}$ can increase by a factor
of 2.
%%%%%%%%%%%%%%%%%%%%%%%%%%%%%%%%%%%%%%%%%%%%%%%%%%%%%%%%%%%%%%%%

%%%%%%%%%%%%%%%%%%%%%%%%%%%%%%%%%%%%%%%%%%%%%%%%%%%%%%%%%%%%%%
\begin{figure}
\centering
\epsfxsize=86mm
\epsffile[80 200 555 680]{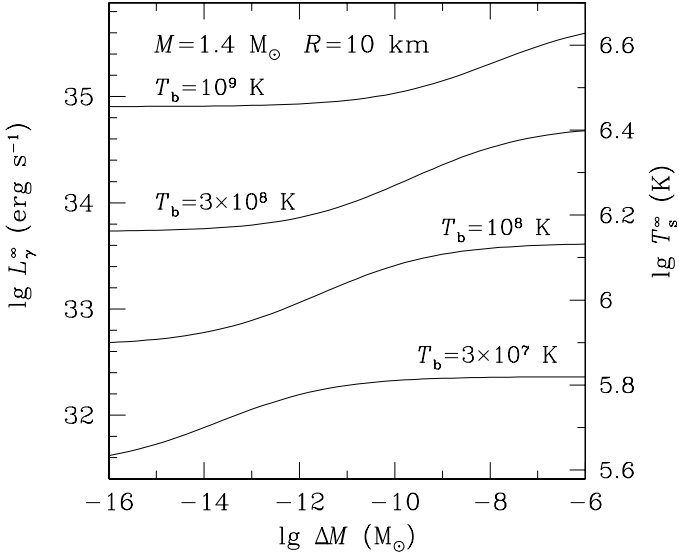}
\caption{
Photon surface luminosity (redshifted as detected by a distant observer,
left vertical axis) or redshifted effective surface temperature
(right vertical axis)
of a ``canonical'' NS model
($M=1.4 \, {\rm M}_\odot$, $R=10$ km) 
for four values of the ``internal'' temperature $T_{\rm b}$
(at the bottom of the heat blanketing layer,
$\rho_{\rm b}=10^{10}$ g cm$^{-3}$) versus
the mass of accreted material, $\Delta M$.
}
\label{fig6}
\end{figure}
%%%%%%%%%%%%%%%%%%%%%%%%%%%%%%%%%%%%%%%%%%%%%%%%%%%%%%%%%%%%%%%

%%%%%%%%%%%%%%%%%%%%%%%%%%%%%%%%%%%%%%%%%%%%%%%%%%%%%%%%%%%%%%
\begin{figure}
\centering
\epsfxsize=86mm
\epsffile[80 200 555 680]{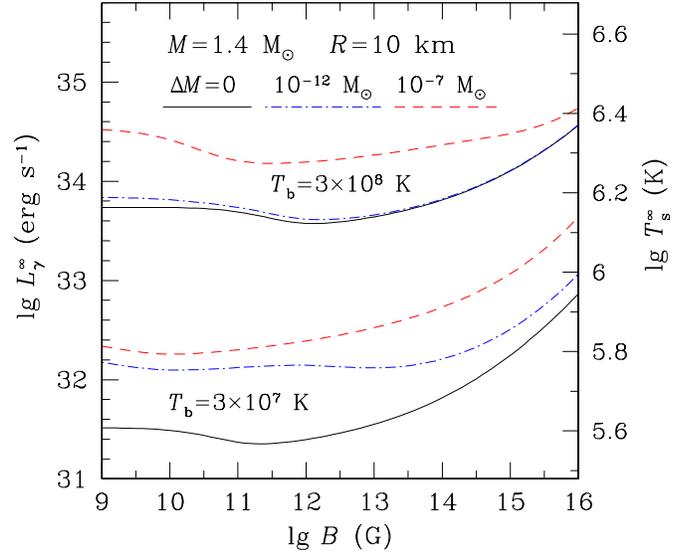}
\caption{
Redshifted photon surface luminosity (left vertical axis)
and mean effective temperature (right vertical axis) of
a ``canonical'' NS with a dipole magnetic field, for
two values of $T_{\rm b}$ and three models of the 
heat-blanketing envelope (accreted mass $\Delta M=0$,
$10^{-12} \, {\rm M}_\odot$, and $10^{-7} \, {\rm M}_\odot$)
versus magnetic field strength $B$ at the magnetic pole.
}
\label{fig7}
\end{figure}
%%%%%%%%%%%%%%%%%%%%%%%%%%%%%%%%%%%%%%%%%%%%%%%%%%%%%%%%%%%%%%%

%%%%%%% 4.4 %%%%%%%%%%%%%%%%%%%%%%%%%%%%%%%%%%%%%%%%%%%%%%%%%%%%%
\subsection{Surface magnetic field}
\label{sect-magnetic}
%%%%%%%%%%%%%%%%%%%%%%%%%%%%%%%%%%%%%%%%%%%%%%%%%%%%%%%%%%%%%%%%%

The effects of strong magnetic field
on the heat conduction through the blanketing NS envelope
have been reviewed, for instance, by
Ventura \& Potekhin (\cite{vp01}).
These effects depend on $\rho$, $T$, and
the field strength $B$. One can distinguish 
(i) {\it classical effects} of electron {\it Larmor
rotation}, and (ii) the {\it effects of 
quantization} of electron orbits across the field
lines (Landau states).

The classical effects require lower magnetic fields.
They become pronounced at
$\omega_{\rm g} \tau\ga 1$,
where $\omega_{\rm g}$ is the electron gyrofrequency, 
and $\tau$ is the effective electron relaxation time.
Then the conduction becomes anisotropic.
The electron thermal conductivity along the field lines
remains equal to its non-magnetic value $\kappa_0$, while the conductivity
in the perpendicular direction becomes 
$\kappa_0/[1+(\omega_{\rm g}\tau)^2]$.

The conditions of electron quantization are determined by
characteristic density $\rho_B$ and temperature $T_B$,
\begin{eqnarray}
   \rho_B &\approx&
   7.045\times10^3 \,(A/Z)
       \,B_{12}^{3/2}\gcc,
\label{rho_B}
\\
  T_B &\approx& {1.343\times10^8\, B_{12}
  \over \sqrt{1+1.018 (\rho_6 Z/A)^{2/3}}}{\rm~K},
\label{T_B}
\end{eqnarray}
where $Z$ is the mean charge number of atomic nuclei,
$A$ is the number of nucleons per a nucleus,
$\rho_6=\rho/10^6\gcc$, and $B_{12}=B/10^{12}$ G.
The magnetic field remains non-quantizing as long as $T \gg T_B$.

If $\rho\ga\rho_B$ and $T \la T_B$, then the magnetic
field acts as {\it weakly quantizing}. In this regime,
the conductivities exhibit de Haas -- van Alphen oscillations
around their classical values.
The field becomes {\it strongly quantizing} at
$\rho \la \rho_B$ and $T \la T_B$; it substantially modifies 
all components of the conductivity tensors well as 
the equation of state.

The thermal structure of the NS envelopes composed of iron
at $B$ up to $10^{16}$ G
was studied, e.g., by Potekhin \& Yakovlev (\cite{py01}).
Consider, for instance, the dipole field. Near the magnetic pole,
the heat is carried away from the NS interior 
along the field lines. The electron-quantization effects 
amplify this conduction and make the polar regions
of the heat-blanketing envelope more heat-transparent,
increasing the local effective surface temperature for a given
$T_{\rm b}$. On the contrary,
near the magnetic equator the heat propagates
across the magnetic field. If $\omega_{\rm g}\tau\gg1$,
the classical Larmor-rotation effects strongly reduce
the corresponding transverse conductivity. 
The equatorial regions become
less heat transparent, which lowers the local effective
temperature for a given $T_{\rm b}$.
In an arbitrary element
of the envelope, the heat conduction is affected
by both (longitudinal and transverse) conductivities.
The NS photon luminosity is obtained by integration
of the local radiated flux over the entire NS surface
-- see Potekhin \& Yakovlev (\cite{py01}) and references therein
for the case of the envelopes composed of iron. 
Here, we supplement these results with new results obtained
for partly accreted envelopes with magnetic fields.
The details of the new calculations 
will be presented elsewhere.
Figure \ref{fig7} displays the photon luminosity
versus $B$ for two selected values of $T_{\rm b}$
and three selected values of $\Delta M$.
The magnetic field affects the luminosity at
$B \ga 3 \times 10^{10}$ G. In the range of $B$
up to $B \sim 3 \times 10^{13}$ G, the equatorial
decrease of the heat transport dominates, and the
NS luminosity is lower than at $B=0$. For higher $B$,
the polar increase of the heat transport becomes
more important, and the magnetic field increases
the photon luminosity.
The variation of the photon luminosity by the
magnetic field $B \la 10^{15}$ G does not exceed
a factor of 30.

The joint effect of the accreted envelope and 
the magnetic field is demonstrated by the dot-dashed and dashed lines.
As in the non-magnetic case, the accreted material makes the 
envelope more heat-transparent, thus increasing the luminosity
at given $T_{\rm b}$. Therefore, at $B\sim10^{10}$--$10^{13}$ G,
the magnetic field and the accreted envelope
affect the thermal insulation
in opposite directions. At higher field strengths,
both effects increase the luminosity.
However, as evident from Fig.\ \ref{fig7}, 
the dependence of this increase on $B$ and $\Delta M$ 
is complicated. In particular, at $B \ga 10^{14}$ G,
the effect of the
accreted envelope is generally
weaker than in the non-magnetic case. 
%%%%%%%%%%%%%%%%%%%%%%%%%%%%%%%%%%%%%%%%%%%%%%%%%%%%%%%%%%%%%%%%%%%%%%%

%%%%%%%%%%%%%%%%%%%%%%%%%%%%%%%%%%%%%%%%%%%%%%%%%%%%%%%%
%%  Sect.\ 4.5.%%%%%%%%%%%%%%%%%%%%%%%%%%%%%%%%%%%%%%%%%%
%\subsection{The strongest cooling regulators}
%%%%%%%%%%%%%%%%%%%%%%%%%%%%%%%%%%%%%%%%%%%%%%%%%%%%%%%%%
%
%Comparing the efficiency of the cooling regulators
%described above we may conclude that the strongest
%effect on the cooling is produced (i) by the open
%direct Urca in the NS kernel; and (ii) by
%the influence of nucleon superfluidity on neutrino
%emission from the NS core.
%%%%%%%%%%%%%%%%%%%%%%%%%%%%%%%%%%%%%%%%%%%%%%%%%%%%%%%

%%%%%%%%%%%%%%%%%%%%%%%%%%%%% Sect. 5 %%%%%%%%%%%%%%%%%%%%%%%%%
\section{Observational data}
\label{sect5}
%%%%%%%%%%%%%%%%%%%%%%%%%%%%%%%%%%%%%%%%%%%%%%%%%%%%%%%%%%%%%%%

We will confront theoretical cooling curves with the results
of observations of thermal emission from eight middle-aged isolated
NSs. In discussing the observations and their theoretical
interpretation we closely follow the consideration in KYG.

The observational data are the same as KYG.
They are summarized in Table 3 and displayed
in Figs.\ 8--13.
% Table 3 contains the references to
% the sources where the data were taken from.
The three youngest objects
(RX J0822--43, 1E 1207--52, and RX J0002+62) are radio-quiet
NSs in supernova remnants. The oldest object, RX J1856--3754,
is also a radio-quiet NS. The other objects, Vela, PSR 0656+14,
Geminga, and PSR 1055--52, are observed as radio pulsars.
The NS ages are either pulsar spindown ages or the estimated
supernova ages. The age of RX J1856--3754 was estimated
by Walter (\cite{walter01}) from the kinematical data
(by identifying a possible companion
in the binary system
existed before the supernova explosion). 
We use the value $t=9 \times 10^5$ yr
mentioned in the subsequent publication by Pons et al.\
(\cite{ponsetal02}).

For the four youngest sources, the effective surface temperatures
$T_{\rm s}^\infty$
are obtained from the observed X-ray spectra using
hydrogen atmosphere models. Such models are more consistent 
(e.g., Pavlov \& Zavlin \cite{pz02}) with other
information on these sources (distances, hydrogen column
densities, inferred NS radii, etc.) than the blackbody model of
NS emission. On the contrary, for the next three sources we present
the values of $T_{\rm s}^\infty$ inferred using the blackbody spectrum
because the blackbody model is more consistent for these sources.
Finally, for RX J1856--3754 we adopt the values inferred
using the analytic fit with Si-ash atmosphere model of Pons et al.\
(\cite{ponsetal02}). We expect that the large error bar of $T_{\rm s}^\infty$
provided by this model reflects poor understanding
of thermal emission from this source
(e.g., Pons et al.\ \cite{ponsetal02}, Burwitz et al.\ \cite{burwitzetal01},
G\"ansicke et al.\ \cite{gbr01}, Kaplan et al.\ \cite{kva01},
Pavlov et al.\ \cite{pavlovetal02}).

%%%%%%%%%%%%%%%%%%%%%%%%%%%%%%%%% Sect. 6 %%%%%%%%%%%%%%%%%%%%%%%%%%%%%%%%%%%
\section{Theory and observations}
\label{sect6}
%%%%%%%%%%%%%%%%%%%%%%%%%%%%%%%%%%%%%%%%%%%%%%%%%%%%%%%%%%%%%%%%%%%%%%%%%%%%%

%%%%%%%%%%%%%%%%%%%%%%%%%%%%%%%%% Sect. 6.1 %%%%%%%%%%%%%%%%%%%%%%%%%%%%%%%%%
\subsection{Non-superfluid neutron stars}
\label{sect6-1}
%%%%%%%%%%%%%%%%%%%%%%%%%%%%%%%%%%%%%%%%%%%%%%%%%%%%%%%%%%%%%%%%%%%%%%%%%%%%%

There is a large scatter of observational limits on $T_{\rm s}^\infty$
for the eight sources. Three sources, the youngest RX J0822--43,
and two oldest, PSR 1055--52 and RX J1856--3754,  seem to be hot for their
ages, while the other ones, especially Vela and Geminga, look much colder.
We can interpret all observational data
with the cooling curves using the fixed (the same) EOS and
models of the critical temperatures $T_{\rm c}(\rho)$ (Sect.\ 3)
for all objects.
The results are presented in
Figs.\ \ref{fig8}--\ref{fig13}.

In the {\it absence of any superfluidity}
there are {\it two} well-known, distinctly different
cooling regimes, {\it slow} and {\it fast} cooling.
The slowly cooling NSs have low masses,
$M< M_{\rm D}$. They mainly loose
their energy via neutrino emission in modified
Urca processes. For given EOSs in the NS core, the cooling
curves of middle-aged NSs are almost the same
for all $M$ from about ${\rm M}_\odot$
to $M_{\rm D}$ (e.g., Page \& Applegate \cite{pa92},
Gnedin et al.\ \cite{gyp01}), being not very
sensitive to EOS.
The fast cooling occurs
via a very powerful direct Urca process
(Lattimer et al.\ \cite{lpph91})
if $M > M_{\rm D}$ + 
%%$0.01 \, {\rm M}_\odot$
$0.003 \, {\rm M}_\odot$.
The cooling curves are again not too sensitive to the mass and EOS.
The middle-aged rapidly cooling NSs are much colder
than the slowly cooling ones.
Two examples, 1.35 and 1.5 M$_\odot$
non-superfluid NSs (EOS A),
are displayed
in Fig.\ \ref{fig8}.
The transition from the slow to fast cooling takes
place in a very narrow mass range,
$M_{\rm D}< M \la M_{\rm D}$+
%%$0.01 \, {\rm M}_\odot$.
$0.003 \, {\rm M}_\odot$.
In order to explain these observational limits with
the non-superfluid NS models one should make an unlikely assumption
that at least the masses of the Vela and Geminga pulsars
fall in that narrow mass range.
Thus, several sources exhibit
the intermediate case between the slow and fast cooling.
In the absence of superfluidity, this is highly unlikely.

%%%%%%%%%%%%%%%%%%%%%%%%%%%%%%% Sect. 6.2 %%%%%%%%%%%%%%%%%%%%%%%%%%
\subsection{Proton superfluidity and the three types of cooling neutron stars}
\label{sect6-2}
%%%%%%%%%%%%%%%%%%%%%%%%%%%%%%%%%%%%%%%%%%%%%%%%%%%%%%%%%%%%%%%%%%%%

The next step is to interpret the observations
(Fig.\ \ref{fig9}--\ref{fig13}) by cooling of superfluid NSs.
It turns out that various superfluids
affect NS cooling in different ways.
Our main assumptions would be that
the {\it proton superfluidity
is rather strong} at $\rho \la \rho_{\rm D}$, while
the $^3$P$_2$
{\it neutron superfluidity is rather weak} (Sect.\ 6.4).
We start with 
the effects of proton superfluidity.

%%%%%%%%%%%%%%%%%%%%%%%%%%%%%%%%%%%%%%%%%%%%%%%%%%%%%%%%%%%%%%
\begin{figure}
\centering
\epsfxsize=86mm
\epsffile[60 200 550 680]{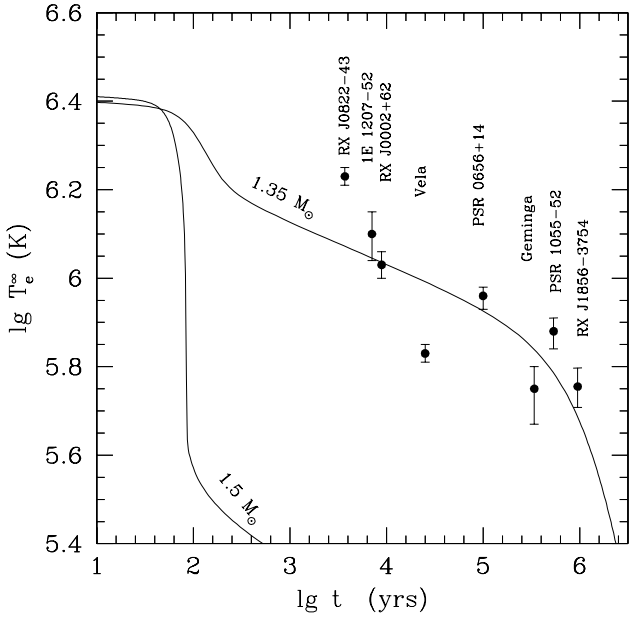}
\caption{
Observational limits on surface temperatures of eight
NSs (Table 3) compared with cooling curves
of non-superfluid NSs (EOS A) of $M=$
1.35 and 1.5 ${\rm M}_\odot$.
}
\label{fig8}
\end{figure}
%%%%%%%%%%%%%%%%%%%%%%%%%%%%%%%%%%%%%%%%%%%%%%%%%%%%%%%%%%%%%%%%%%%%

%%%%%%%%%%%%%%%%%%%%%%%%%%%%%%%%%%%%%%%%%%%%%%%%%%%%%%%%%%%%%%
\begin{figure}
\centering
\epsfxsize=86mm
\epsffile[20 143 590 720]{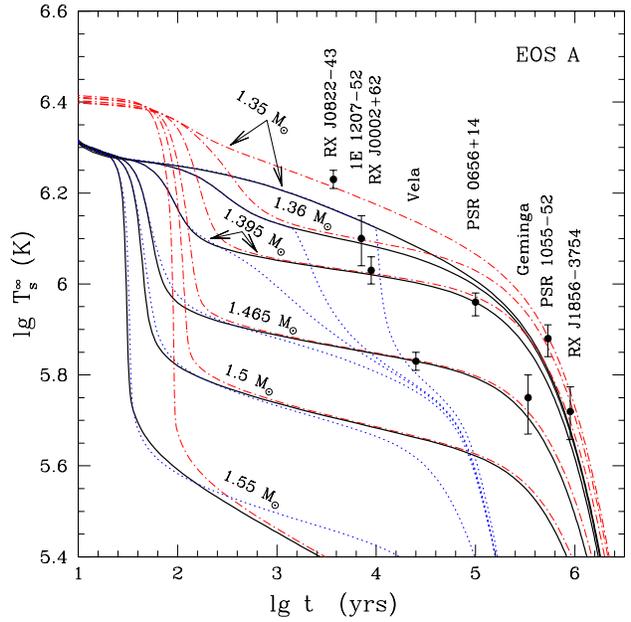}
\caption{
Observational limits on surface temperatures of eight
NSs compared with cooling curves
of NSs (EOS A) with masses from
1.35 to 1.55 ${\rm M}_\odot$ (after KYG).
Dot-and-dashed curves are obtained
including proton superfluidity 1p alone.
Solid curves include, in addition,
model 1ns of neutron superfluidity.
Dotted lines also take into account
the effect of neutron superfluidity 1nt.
}
\label{fig9}
\end{figure}
%%%%%%%%%%%%%%%%%%%%%%%%%%%%%%%%%%%%%%%%%%%%%%%%%%%%%%%%%%%%%%%%%%%%

%%%%%%%%%%%%%%%%%%%%%%%%%%%%%%%%%%%%%%%%%%%%%%%%%%%%%%%%%%%%%%
\begin{figure}
\centering
\epsfxsize=86mm
\epsffile[20 143 590 720]{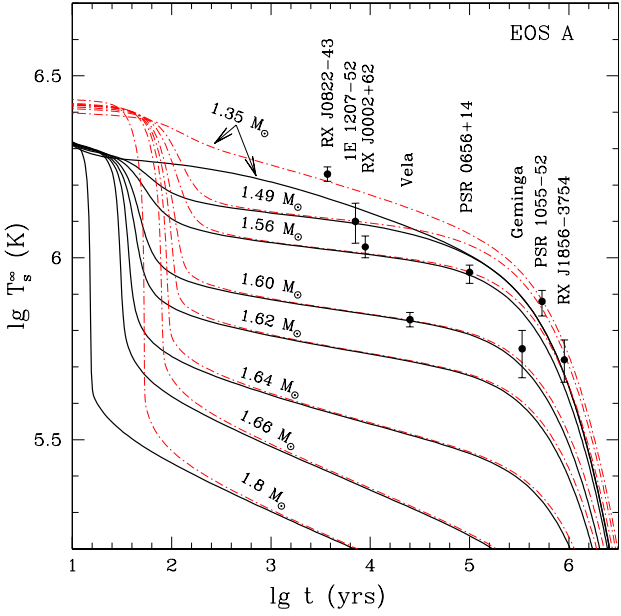}
\caption{
Observational limits on surface temperatures of
NSs compared with cooling curves
of NSs (EOS A) with several masses $M$ in the presence of
proton superfluidity 2p (from KYG).
Dot-and-dashed curves are obtained
assuming non-superfluid neutrons.
Solid curves include, in addition,
model 1ns of neutron superfluidity.
$^3$P$_2$ neutron pairing is neglected.
}
\label{fig10}
\end{figure}
%%%%%%%%%%%%%%%%%%%%%%%%%%%%%%%%%%%%%%%%%%%%%%%%%%%%%%%%%%%%%%%%%%%%

%%%%%%%%%%%%%%%%%%%%%%%%%%%%%%%%%%%%%%%%%%%%%%%%%%%%%%%%%%%%%%
\begin{figure}
\centering
\epsfxsize=86mm
\epsffile[20 143 590 720]{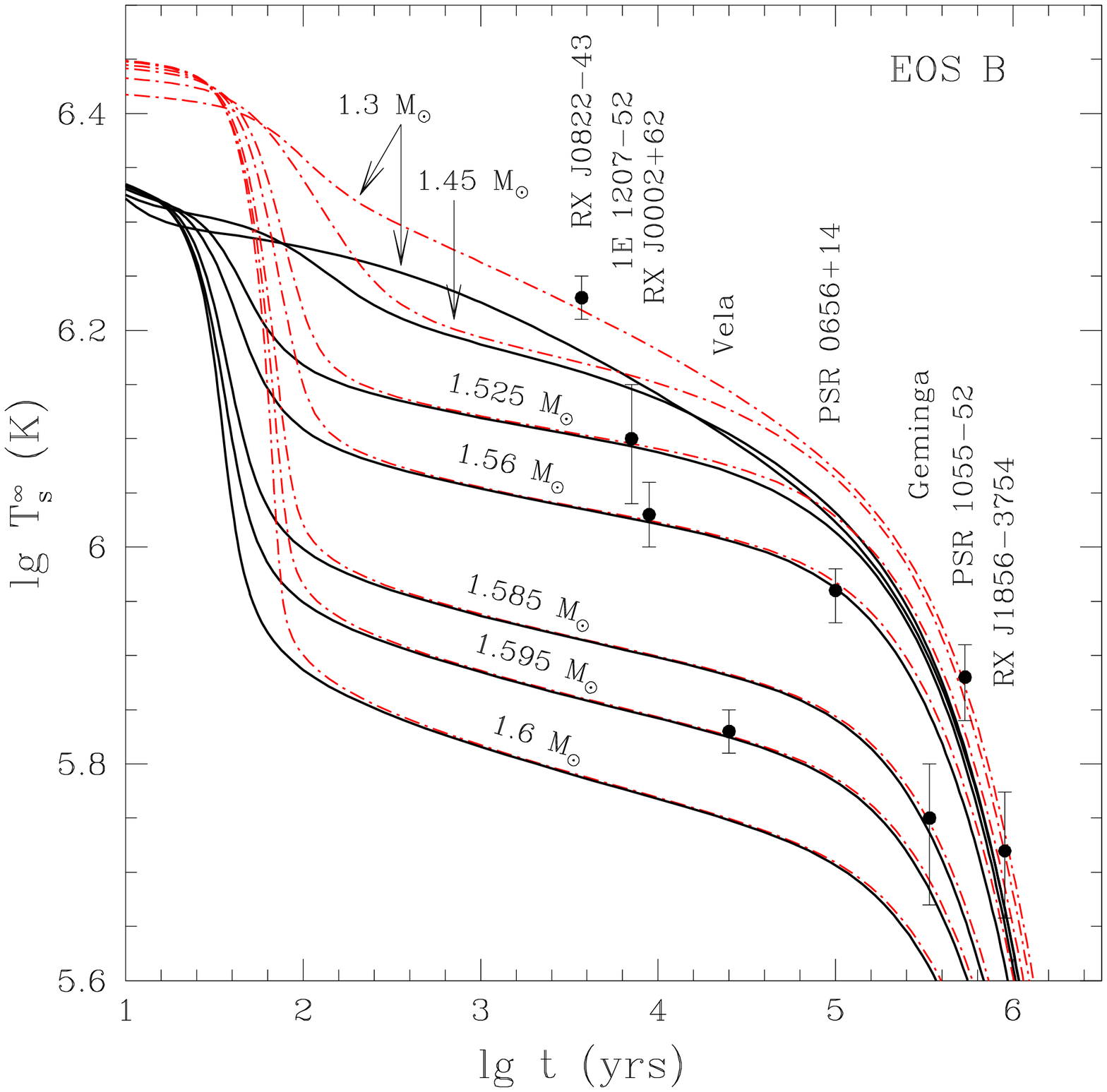}
\caption{
Observational limits on surface temperatures of
NSs compared with cooling curves
of NSs (EOS B) with masses from
1.3 to 1.6 ${\rm M}_\odot$ (from KYG).
Dot-and-dashed curves are obtained
using model 3p of proton superfluidity.
Solid curves include, in addition,
model 1ns of neutron superfluidity.
}
\label{fig11}
\end{figure}
%%%%%%%%%%%%%%%%%%%%%%%%%%%%%%%%%%%%%%%%%%%%%%%%%%%%%%%%%%%%%%%%%%%%

The dot-and-dashed cooling curves in Figs.\ \ref{fig9}--\ref{fig11}
are computed
assuming the proton superfluidity alone.
The proton pairing 1p is adopted
in Fig.\ \ref{fig9}, 2p in Fig.\ \ref{fig10},
and 3p in Fig.\ \ref{fig11}. EOS A is used in the models
in Figs.\ \ref{fig9} and \ref{fig10}, and EOS B in Fig.\ \ref{fig11}.

As seen from Figs.\ \ref{fig9}--\ref{fig11},
the proton superfluidity
leads to the {\it three} cooling regimes
(instead of two in non-superfluid NSs):
{\it slow, moderate}, and {\it fast}.
They reflect three neutrino emission regimes
from the NS cores discussed in Sect.\ 4.2 (Fig.\ \ref{fig3}).
Accordingly, there are
{\it three types} of cooling NSs with distinctly
different properties.

(I) {\it Low-mass, slowly cooling} NSs.
Their central densities $\rho_{\rm c}$
and masses $M=M(\rho_{\rm c})$  obey
the inequalities
\begin{equation}
  \rho_{\rm c}\la \rho_{\rm I}, \quad
  M \la M_{\rm I} = M(\rho_{\rm I}),
\label{MI}
\end{equation}
%
%%the threshold value
where
$\rho_{\rm I}$, introduced in Sect.\ 4.2,
is discussed below in more detail.

(II) {\it Medium-mass, moderately cooling} NSs, with
\begin{equation}
   \rho_{\rm I} \la \rho_{\rm c} \la \rho_{\rm II},
   \quad \quad
   M_{\rm I} \la M \la M_{\rm II}=M(\rho_{\rm II}),
\label{MII}
\end{equation}
where $\rho_{\rm II}$, also
introduced in Sect.\ 4.2, is specified below.

(III) {\it Massive, rapidly cooling} NSs,
\begin{equation}
  \rho_{\rm II} \la \rho_{\rm c} \leq \rho_{\rm cmax}, \quad
   M_{\rm II} \la M \leq M_{\rm max},
\label{MIII}
\end{equation}
where $\rho_{\rm cmax}$ and $M_{\rm max}$ refer to
the maximum-mass configuration (Table 1).

The threshold values of $\rho_{\rm I,II}$ and
$M_{\rm I,II}$ depend on a proton superfluidity model
$T_{\rm cp}(\rho)$,
EOS in the NS core, and on
a NS age. The main properties of regimes I, II, and III
are as follows.

(I) {\it The slowly cooling NSs}
are those where the direct Urca
process is either forbidden by momentum conservation
($\rho_{\rm c} \leq \rho_{\rm D}$) 
%%%%%%%%%%%%%%%%%%%%
%(Lattimer et al.\ \cite{lpph91})
%%%%%%%%%%%%%%%%%%%%
or greatly suppressed by the strong proton superfluidity
(Fig.\ \ref{fig3}).

In particular, the
cooling is slow for $\rho_{\rm c} < \rho_{\rm D}$
and $M < M_{\rm D}$ in the absence of proton superfluidity.
This is the well-known {\it ordinary slow cooling} 
which is mainly regulated by
the neutrino emission produced by the
modified Urca process (Sect.\ 6.1).
However, for the conditions displayed in
Figs.\ \ref{fig9}--\ref{fig11},
the proton superfluidity in regime I is so strong
that it almost switches off both,
the modified Urca process everywhere in the NS core
and the direct Urca process at $\rho > \rho_{\rm D}$.
Then the neutrino emission
is mainly produced by neutrino bremsstrahlung in
neutron-neutron scattering (Sect.\ 4.2) unaffected by
the neutron superfluidity in the NS core
(that is assumed to be weak).
The bremsstrahlung is less efficient than
the modified Urca process and leads
to a slower cooling than
in a non-superfluid NS. It can be referred to
as the {\it very slow} cooling.

The analysis shows (KYG) that,
for the selected models,
the regime of very slow cooling holds as long as
the proton critical temperature in the NS center is higher
than a threshold value:
\begin{equation}
      T_{\rm cp}(\rho_{\rm c}) \ga T_{\rm cp}^{\rm (I)}(\rho_{\rm c}).
\label{TcpI}
\end{equation}
Comparing the neutrino emissivities of the indicated
reactions (e.g., Yakovlev et al.\ \cite{yls99}), KYG
obtained the simple estimates:
$T_{\rm cp}^{\rm (I)}(\rho) \sim 5.5 \, T$ for
$\rho \leq \rho_{\rm D}$, and
$T_{\rm cp}^{\rm (I)}(\rho) \sim 17 \, T$ for
any $\rho$ which is several per cent higher than $\rho_{\rm D}$,
where $T$ is the internal NS temperature.
There is a continuous transition (Fig.\ 9 in KYG) 
between the presented values of $T_{\rm cp}^{\rm (I)}(\rho)$
in the narrow density range near $\rho_{\rm D}$.

To make the analysis less abstract we notice that
%%%%%%%%%%%%%%%
%$T \sim 5.5 \times 10^8$ K in a very slowly cooling NS
%at $t \sim 4 \times 10^3$ yr,
%$T \sim 4 \times 10^8$ K at $t \sim 2.5 \times 10^4$ yr,
%and $T \sim 1.5 \times 10^8$ K at $t \sim 4 \times 10^5$ yr.
%%%%%%%%%%%%%%%%
in a very slowly cooling NS 
the internal temperature is $T \sim 5.5 \times 10^8$, 
$4 \times 10^8$, and $1.5 \times 10^8$~K at 
$t \sim 4 \times 10^3$, $2.5 \times 10^4$, and $4 \times 10^5$~yr,
respectively.  

Now we can explicitly specify the maximum
central densities $\rho_{\rm I}$ and masses $M_{\rm I}$
of very slowly cooling NSs in Eq.\ (\ref{MI})
for the cases of study, in which
$\rho_{\rm I} \ga \rho_{\rm D}$ 
($M_{\rm I} \ga M_{\rm D}$).
If $\rho_{\rm c} > \rho_{\rm D}$, then
we can define $\rho_{\rm I}$ as the density value
on the decreasing, high-density slope of
$T_{\rm cp}(\rho)$ (Figs.\ \ref{fig1} and \ref{fig2})
which corresponds to
$T_{\rm cp}(\rho_{\rm I}) =
T_{\rm cp}^{\rm (I)}(\rho_{\rm I})$.
It gives the central density of the NS
and $M_{\rm I}=M(\rho_{\rm I})$.

For the conditions displayed
in Figs.\ \ref{fig9}--\ref{fig11}, the
cooling curves (dot-and-dashed lines) of all
low-mass NSs with any proton superfluidity
(1p, 2p, or 3p) are very similar.
For instance,
the $1.35 \, {\rm M}_\odot$ curve in Fig.\ \ref{fig3}
is plotted just as an example; all the curves are almost
identical in the mass range ${\rm M}_\odot \la M < M_{\rm I}$.
Moreover, the  
cooling
curves are not too sensitive to EOS
and are {\it insensitive to the exact values of the
proton critical
temperature} $T_{\rm cp}$,
as long as the inequality (\ref{TcpI}) holds.
These curves are noticeably higher than the curves
which describe the ordinary slow cooling in the
absence of superfluidity (e.g., Fig.\ \ref{fig8}).
%%%%%%%%%%%%%%%%%%%%%%%%
%They give the highest surface temperatures $T_{\rm s}^\infty$
%of NSs with strong proton superfluidity
%(without any internal reheating or 
%the accreted or magnetized envelopes, Sect.\ 6.3).
%%%%%%%%%%%%%%%%%%%%%%%%
It is remarkable that they seem
to be almost model-independent. 

For the conditions in Figs.\ \ref{fig9}--\ref{fig11},
the three relatively hot sources,
RX J0822--43, PSR 1055--52, and RX J1856--3754, can be treated
as these very-slow-cooling (low-mass)
models.

(II) {\it The moderately cooling stars}
are the NSs which possess central kernels
where the direct Urca process is allowed but
moderately suppressed by proton superfluidity.
The existence of a representative class of these NSs
is solely due to proton superfluidity.

According to KYG, for the selected cooling models,
the proton critical temperature
in the center of a medium-mass NS should roughly satisfy the inequality
\begin{equation}
   T_{\rm cp}^{\rm (II)}(\rho_{\rm c})
   \la T_{\rm cp}(\rho_{\rm c}) \la
   T_{\rm cp}^{\rm (I)}(\rho_{\rm c}),
\label{TcpII}
\end{equation}
with $T_{\rm cp}^{\rm (II)} \sim 3 \, T$.
Thus, one can find $\rho_{\rm II}$ which corresponds
to $T_{\rm cp}(\rho_{\rm II}) = T_{\rm cp}^{\rm (II)}(\rho_{\rm II})$
and determines
$M_{\rm II}$, the maximum mass of moderately
cooling NSs in Eq.\ (\ref{MII}). 
%%%%%%%%%%%%%%%%%%%%%%%%%%%%%%%%%%%%%
%The values of
%$T_{\rm cp}^{\rm (II)}$ depend mainly
%on a given EOS and slightly
%on a NS age (KYG). The ranges of mass
%and density in Eq.\ (\ref{MII}) depend
%also on a model of $T_{\rm cp}(\rho)$.
%%%%%%%%%%%%%%%%%%%%%%%%%%%%%%%%%%%%%%

The surface temperatures of these medium-mass
NSs
are governed by proton superfluidity in the NS
central kernels, $\rho \ga \rho_{\rm I}$.
One can observe (Figs.\ \ref{fig9}--\ref{fig11})
a steady decrease of  
the surface temperatures with increasing $M$.
Fixing the proton superfluidity and EOS,
one can determine (Kaminker et al.\ \cite{khy01} and
Yakovlev et al.\ \cite{ykg01}) the masses of
moderately cooling NS, which means
``weighing'' NSs.
In this fashion one can weigh five isolated NSs
(1E 1207--52, RX J0002+62, Vela, PSR 0656+14, and
Geminga) as shown in Figs.\ \ref{fig9}--\ref{fig11}.
For instance, adopting
EOS A and proton superfluid 1p (Fig.\ \ref{fig9})
KYG obtain the masses in the range from
$\approx 1.36 \, {\rm M}_\odot$ (for 1E 1207--52) to
$\approx 1.465 \, {\rm M}_\odot$ (for Vela and Geminga).
For EOS A and proton superfluid 2p (Fig.\ \ref{fig10})
KYG obtain higher masses of the same sources.
Obviously, the properties of moderately cooling NSs
are {\it extremely sensitive} to the decreasing
slope of $T_{\rm cp}(\rho)$ in the density
range from $\rho_{\rm I}$ to $\rho_{\rm II}$
(and insensitive to the details
of $T_{\rm cp}(\rho)$ outside this range).

(III) {\it Massive} NSs show
{\it fast} cooling similar to the fast cooling of
non-superfluid NSs. These stars have central kernels where
the direct Urca process is either unaffected or weakly suppressed
by the proton superfluidity. In such kernels,
$T_{\rm cp}(\rho) \la T_{\rm cp}^{\rm (II)}$.
The central densities and masses of these NSs lie in the range
given by Eq.\ (\ref{MIII}).
Their thermal evolution
is not very sensitive to
the model of $T_{\rm cp}(\rho)$ and to EOS
in the stellar core. Note that if $\rho_{\rm cmax}< \rho_{\rm II}$,
the rapidly cooling NSs do not exist.
In the frame of this interpretation,
no NS observed so far can be assigned to this class.

%%%%%%%%%%%%%%%%%%%%%%%%%%%%%%% Sect. 6.3 %%%%%%%%%%%%%%%%%%%%%%%%%%
\subsection{Crustal superfluidity and slow cooling}
\label{sect6-3}
%%%%%%%%%%%%%%%%%%%%%%%%%%%%%%%%%%%%%%%%%%%%%%%%%%%%%%%%%%%%%%%%%%%%

At the next step, we retain proton superfluidity
and add $^1$S$_0$ neutron superfluidity 1ns in the NS crust.
These models are shown by the solid curves
in Figs.\ \ref{fig9}--\ref{fig11}.
For the moderately or rapidly cooling middle-aged NSs,
they are fairly close to
the dot-and-dashed curves. This is quite expected
(e.g., Gnedin et al.\ \cite{gyp01}): the $^1$S$_0$
neutron superfluidity is mainly located in the NS crust which is much less
massive than the NS core. Thus,
the crustal superfluidity does not affect noticeably
% cooling in the isothermal regime and
the proposed interpretation
of 1E 1207--43, RX J0002+62, Vela, PSR 0656+14, and Geminga
in terms of moderately cooling NSs.

However, as pointed out by Yakovlev et al.\ (\cite{ykg01}),
this crustal superfluidity strongly affects the slow cooling
of low-mass NSs, and
the effects are twofold. First, at $t \la 3 \times 10^5$ yr
the neutrino luminosity due to $^1$S$_0$ pairing of neutrons
may dominate
the sufficiently low neutrino luminosity
of the stellar core.
Second, at $t \ga 10^5$ yr the $^1$S$_0$ neutron superfluidity
reduces the heat capacity of the crust. Both effects
accelerate NS cooling and decrease $T_{\rm s}^\infty$
(Figs.\ \ref{fig9}--\ref{fig11})
violating the interpretation of the three sufficiently hot sources,
RX J0822--43, PSR 1055--52, and RX J1856--3754.
The interpretation of RX J1856--3754
is affected to a lesser extent,
as a consequence of the rather large error bar of $T_{\rm s}^\infty$
for this source  % RX J1856--3754
(Sect.\ 5).

\begin{figure}
\centering
\epsfxsize=86mm
\epsffile[20 143 590 720]{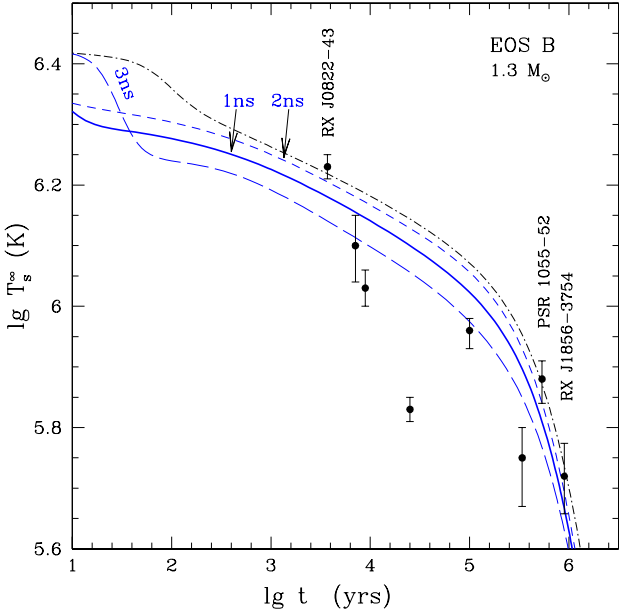}
\caption{
Cooling curves of $1.3 \, {\rm M}_\odot$ NS model (EOS B)
versus observations of RX J0822--43, PSR 1055--52,
and RX J1856--3754 (after KYG). Dot-and-dashed curve:
proton superfluidity 3p in the NS core. Solid, short-dashed,
and long-dashed curves include, in addition,
models 1ns, 2ns, and 3ns of crustal neutron superfluidity, respectively.
Thick solid line is the same as
in Fig.\ \ref{fig11}. 
}
\label{fig12}
\end{figure}

\begin{figure}
\centering
\epsfxsize=86mm
\epsffile[20 143 590 720]{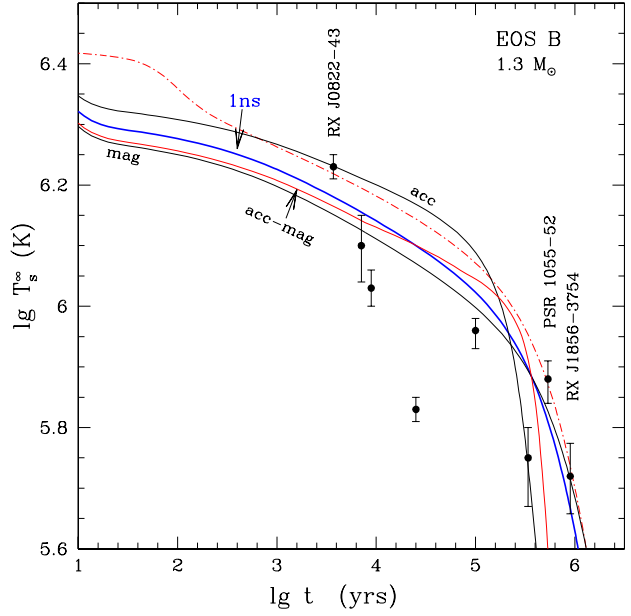}
\caption{
Cooling curves of $1.3 \, {\rm M}_\odot$ NS model
versus observations of RX J0822--43, PSR 1055--52,
and RX J1856--3754. Dot-and-dashed curve:
proton superfluidity 3p in the NS core. 
Solid curves include, in addition,
model 1ns of crustal neutron superfluidity.
Thick solid line is the same as
in Figs.\ \ref{fig11} and \ref{fig12}. 
Thin solid curve {\it acc} is calculated
assuming the presence of $2 \times 10^{-11}\, {\rm M}_\odot$ of
hydrogen on the NS surface. Thin solid curve
{\it mag} is obtained assuming the iron surface and dipole
magnetic field ($10^{12}$ G at the
magnetic pole). Thin solid curve {\it acc-mag}
is obtained assuming both, the magnetic field
($10^{12}$ G) and accreted envelope
($2 \times 10^{-11}\, {\rm M}_\odot$).
}
\label{fig13}
\end{figure}

However, the interpretation can be rescued by the
appropriate choice of $T_{\rm cns}(\rho)$.
It is sufficient to focus on the interpretation of
RX J0822--43, PSR 1055--52, and RX J1856--3754, as
the very slowly cooling NSs.
For certainty, we take EOS B, $M=1.3 \, {\rm M}_\odot$,
and proton superfluid 3p. 
The results are presented in Figs.\ \ref{fig12} and \ref{fig13}.
The dot-and-dashed line is the same as in Fig.\ \ref{fig11} and neglects
the crustal neutron superfluidity.
Thick solid line is also the same as in Fig.\ \ref{fig11}.
It includes an additional effect of
crustal superfluid 1ns and lies below the
observational limits on $T_{\rm s}^\infty$ for the sources in question
(or almost below in case of RX J1856--3754).
To keep the proposed interpretation of the three sources
one must raise the cooling curves calculated including the
crustal superfluidity.
To this aim, one should suppress the neutrino emission associated with
$^1$S$_0$ pairing of neutrons (Fig.\ \ref{fig5}).
Recall that in a middle-aged NS
this emission is mainly
generated (Fig.\ \ref{fig5}) in two relatively narrow layers,
near the neutron drip point and near the crust-boundary interface,
where the local NS temperature $T$ is just below $T_{\rm cns}(\rho)$.
Since the Cooper-pairing
neutrino luminosity is roughly proportional
to the widths of these emitting  
layers,
one can reduce the luminosity by reducing the widths.
This can be done by setting $T_{\rm cns}^{\rm max}$
higher and by making the sharper decrease of
$T_{\rm cns}(\rho)$ 
in the wings.

For example, taking crustal superfluid 2ns
instead of 1ns (Figs.\ \ref{fig4} and \ref{fig5}) one obtains the dashed
cooling curve in Fig.\ \ref{fig12} which comes much closer
to the dot-and-dashed curve than the thick solid curve (model 1ns).
Note that the cooling curves
are insensitive to the details of $T_{\rm cns}(\rho)$
profile near the maximum,  
as long as
$T_{\rm cns}^{\rm max}\ga 5 \times 10^9$ K,
but they are extremely sensitive to the decreasing
slopes of $T_{\rm cns}(\rho)$.
On the other hand, by taking the smoother and lower
$T_{\rm cns}(\rho)$, model 3ns, one obtains
a colder NS than
needed for the interpretation of the observations
(long-dash line in Fig.\ \ref{fig12}).
Therefore, $^1$S$_0$ neutron
superfluidity with maximum $T_{\rm cns}^{\rm max} < 5 \times 10^9$ K
and/or with smoothly decreasing slopes of the $T_{\rm cns}(\rho)$
profile near the crust-core interface and the
neutron drip point {\it violates}
the proposed interpretation of the observational data.

Moreover, the observations
of RX J0822--43, PSR 1055--52, and RX J1856--3754
can be fitted even with the
initial model 1ns of the crustal superfluidity
(Fig.\ \ref{fig13}). The
high surface temperature of RX J0822--43 can be explained
assuming additionally the presence of a low-mass
($2 \times 10^{-11} \, {\rm M}_\odot$) heat-blanketing
surface envelope of hydrogen or helium.
This effect is modeled using
the results of Potekhin et al.\ (\cite{pcy97}) described in Sect.\ 4
(Fig.\ \ref{fig6}).
Light elements
raise $T_{\rm s}^\infty$ at the
neutrino cooling stage (curve {\it acc} in Fig.\ \ref{fig13}).
In order to explain the observations of PSR 1055--52
and RX J1856--3754, one can assume again
model 1ns of crustal superfluidity,
iron surface and the dipole surface magnetic field
($\sim 10^{12}$ G at the magnetic pole;
line {\it mag} in Fig.\ \ref{fig13}).
Such a field makes the NS surface layers
overall less heat-transparent (Fig.\ \ref{fig7}),
rising $T_{\rm s}^\infty$
at $t \ga 3 \times 10^5$ yr.
Note that the dipole field $\ga 3 \times 10^{13}$~G
has the opposite effect, resembling the effect of
the surface envelope of light elements.

To summarize, one can additionally vary cooling
curves by assuming the presence of light elements
%%%%%%%
%and/or 
%%%%%%
and
the magnetic field on the NS surface
(line {\it acc-mag} in Fig.\ \ref{fig13}). 
However, these variations are less pronounced than those due to 
nucleon superfluidity. For instance, one cannot reconcile the cooling curves
with the present observations of PSR 1055--52
assuming model 3ns of the crustal superfluidity
with any surface magnetic field.

%%%%%%%%%%%%%%%%%%%%%%%%%%%%%%% Sect. 6.4 %%%%%%%%%%%%%%%%%%%%%%%%%%
\subsection{$^3${\rm P}$_2$ pairing of neutrons in the NS core}
\label{sect6-4}
%%%%%%%%%%%%%%%%%%%%%%%%%%%%%%%%%%%%%%%%%%%%%%%%%%%%%%%%%%%%%%%%%%%%

Now we focus on the $^3$P$_2$ neutron
pairing neglected so far. Its effects are illustrated
in Fig.\ \ref{fig9}, as an example. They
are qualitatively similar for the other cooling
models in Figs.\ \ref{fig10} and \ref{fig11}.
In Fig.\ \ref{fig9} we take the cooling models
obtained including proton superfluidity 1p
and crustal superfluidity 1ns, and add the $^3$P$_2$
neutron superfluidity (model 1nt, Table 2) in the core.
This gives the same
(solid) cooling curves for the young NSs
which have the internal temperatures $T$ above the maximum value of
$T_{\rm cnt}^{\rm max} \approx 3 \times 10^8$ K. However, when $T$
falls below $T_{\rm cnt}^{\rm max}$, one obtains
(dots) a strong acceleration
of the cooling associated with the powerful neutrino emission
due to $^3$P$_2$ neutron pairing (Fig.\ \ref{fig3}).
This emission greatly
complicates the proposed interpretation of
older sources, PSR 0656+14, Geminga, PSR 1055--52, and
RX J1856--3754.
The complication arises for a wide class of $T_{\rm cnt}(\rho)$
profiles with $T_{\rm cnt}^{\rm max}$ from 
$\sim 10^8$ K to $\sim 3 \times 10^9$ K (in this way,
it can be regarded as model-independent).
The Cooper-pairing neutrino emission of neutrons
induces {\it really fast cooling} of older sources even if
their mass is {\it low}, $M < M_{\rm D}$ (Sect.\ 6.2).
To avoid this difficulty one can assume
(Kaminker et al.\ \cite{khy01}) {\it weak} $^3$P$_2$
{\it pairing}, $T_{\rm cnt}(\rho)$, with maximum
$T_{\rm cnt}^{\rm max} < 10^8$ K;
it does not affect the proposed interpretation.

%%%%%%%%%%%%%%%%%%%%%%%%%%%%%%%%%%%%%%%%%%%%%%%%%%%%%%%%%%%%%%
%%%%%%%%%%%%%%%%%%%%%%%%%%%%%%%%%%%%%%%%%%%%%%%%%%%%%%%%%%%%%%%%%%%%

%%%%%%%%%%%%%%%%%%%%%%%%%%%%%%%%%%%%%%%%%%%%%%%%%%%%%%%%%%%%%%
%%%%%%%%%%%%%%%%%%%%%%%%%%%%%%%%%%%%%%%%%%%%%%%%%%%%%%%%%%%%%%%%%%%%

%%%%%%%%%%%%%%%%%%%%%%%%%%%  Section 7  %%%%%%%%%%%%%%%%%%%%%%%%
\section{Summary}
\label{sect-discuss}
%%%%%%%%%%%%%%%%%%%%%%%%%%%%%%%%%%%%%%%%%%%%%%%%%%%%%%%%%%%%%%%%

Following KYG we summarize the effects superfluids on
NS cooling.

(a) Strong proton superfluidity in the NS cores,
combined with the direct Urca process at
$\rho > \rho_{\rm D}$, separates (Sect.\ 6.2) the cooling models into
three distinctly different
types: (I) slowly cooling, low-mass NSs ($M \la M_{\rm I}$); 
(II) moderately cooling, medium-mass NSs ($M_{\rm I} \la M \la M_{\rm II}$);
(III) rapidly cooling, massive NSs ($M \ga M_{\rm II}$).
The regime of moderate cooling
cannot be realized without the proton superfluidity.

(b) Strong proton superfluidity in the NS core is required to interpret
the observational data on the three sources,
RX J0822--43, PSR 1055--52, and
RX J1856--3754,
hot for their ages,   % hotter got their ages,
as the very slowly cooling NSs
(Sects.\ 6.2 and 6.3).
Within this interpretation,
all three sources may have masses
from about ${\rm M}_\odot$ to $M_{\rm I}$;
one cannot determine their masses exactly
or distinguish EOS in the NS core from the cooling models.

(c) Strong proton superfluidity is needed to interpret
observations of the other sources, 1E 1207--52,
RX J0002+62, Vela, PSR 0656+14, and Geminga,
as the medium-mass NSs.
One can ``weigh'' these NSs, i.e., determine their
masses, for a given model of $T_{\rm cp}(\rho)$ and a given EOS.
The weighing is very sensitive to
the decreasing slope of $T_{\rm cp}(\rho)$
in the density range $\rho_{\rm I} \la  \rho  \la  \rho_{\rm II}$ 
(Sect.\ 6.2).

(d) Strong or moderate $^3$P$_2$ neutron superfluidity
in the NS core initiates rapid cooling due to the neutrino emission
resulted from neutron pairing. This would invalidate the proposed
interpretation of the old sources like PSR 0656+14,
Geminga, PSR 1055--52, and RX J1856--3754.
One can save the interpretation assuming a weak $^3$P$_2$
neutron superfluidity, $T_{\rm cnt}^{\rm max} < 10^8$ K (Sect.\ 6.4).

(e) $^1$S$_0$ neutron superfluidity in the crust
may initiate a strong Cooper-pairing neutrino emission,
decrease substantially $T_{\rm s}^\infty$
of the slowly cooling NSs, and weaken
the interpretation
of RX J0822--43, PSR 1055--52, and RX J1856--3754 (although it does not
affect significantly the moderate or fast cooling).
The interpretation can be saved by assuming
that the maximum of 
$T_{\rm cns}(\rho)$ is not too small
($T_{\rm cns}^{\rm max} \ga 5 \times 10^9$ K) and the
profile of $T_{\rm cns}(\rho)$ decreases
sharply in the wings (Sect.\ 6.3).

(f) The interpretation of the slowly cooling sources
is sensitive to
the presence of the surface magnetic fields and/or
heat-blanketing surface layer composed
of light elements (Sect.\ 6.3).

(g) No isolated middle-aged NSs
observed so far can be idenfied as a
rapidly cooling NS. In the frame of the proposed models, these NSs
do not exist
for those EOSs and $T_{\rm cp}(\rho)$ profiles 
for which $M_{\rm max}< M_{\rm II}$.

If the proposed interpretation is correct, 
one can make the following conclusions
on the properties of dense matter in NS interiors.

(i) Strong proton superfluidity is in favor of
a not too large symmetry energy at supranuclear densities 
(Kaminker et al.\ \cite{khy01}).
A very large symmetry energy would mean a high proton fraction
which would suppress proton pairing. On the other hand,
the symmetry energy should be not too small to open the direct Urca
process at $\rho > \rho_{\rm D}$.

(ii) Weak $^3$P$_2$
neutron pairing is in favor of a not too soft EOS
in the NS core (Kaminker et al.\ \cite{khy01}). 
The softness would mean a strong attractive
neutron-neutron interaction and, therefore, strong neutron pairing.

(iii) The adopted features of the crustal neutron superfluidity
are in favor of those microscopic theories which predict
$T_{\rm cns}(\rho)$ profiles with $T_{\rm cns}^{\rm max}
\ga 5 \times 10^9$ K.
This is in line with many microscopic calculations
of the superfluid gaps which include the medium polarization effects in
neutron-neutron
interaction (e.g., Lombardo \& Schulze \cite{ls01}).
However, the reduction of the gap
by the medium polarization should not be too strong, and
the decreasing slope of $T_{\rm cns}(\rho)$
should be rather sharp.
These requirements constrain the microscopic theories.

The proposed interpretation of the observations
relates the inferred NS masses to the superfluid properties of NS interiors.
By varying EOS and the proton critical temperature, 
one can attribute different masses to the same sources.
If, on the other hand, one knew the range of masses
of the cooling middle-aged NSs it would be possible to draw definite
conclusions on the superfluid state of their interiors,
first of all, on the proton critical temperature, $T_{\rm cp}(\rho)$.

The presented analysis may seem too simplified because
it neglects a possible presence of other particles in the NS cores
(muons, hyperons, quarks). It is expected that 
the inclusion of other particles and the effects
of superfluidity of hyperons or quarks will
complicate theoretical analysis but will not
change the basic conclusion on the existence
of the slowly, moderately, and rapidly cooling NSs.

The calculations show that the cooling of middle-aged NSs
with $M < M_{\rm I}$ is sensitive to the density profile
of free neutrons near the crust bottom and
neutron drip point. The presented calculations used only one
model of the free-neutron distribution in the crust, assuming
spherical atomic nuclei at the crust bottom.
It would be interesting to consider the models
of crust matter with non-spherical nuclei
(e.g., Pethick \& Ravenhall \cite{pr95})
and
the effects of superfluidity of nucleons confined in the atomic nuclei
in the NS crust.

The determination of $T_{\rm s}^\infty$
from observational data is a very complicated problem
(as described, e.g., by Pavlov \& Zavlin \cite{pz02}).
It requires very high-quality data and
theoretical models of NS atmospheres.
Thus, the current values of $T_{\rm s}^\infty$ may change
substantially after the forthcoming observations
and new theoretical modeling. These changes may affect
the proposed interpretation of the observational data,
first of all, of RX J0822--43, PSR 1055--52, and
RX J1856--3754. For instance, RX J1856--3754
may have a colder surface ($T_{\rm s}^\infty \sim 0.25$ MK),
than assumed in the above analysis,
with a hot spot (e.g., Pons et al.\ \cite{ponsetal02},
Burwitz et al.\ \cite{burwitzetal01},
G\"ansicke et al.\ \cite{gbr01}). If confirmed,
the lower $T_{\rm s}^\infty$ might be explained by
the effect of $^3$P$_2$ neutron pairing (Fig.\ \ref{fig9}).

The future observations
of the thermal emission from these sources will be
crucial for understanding the superfluid properties
of NS matter.

The proposed theory can explain the existence of NSs
within a broad range of $T_{\rm s}^\infty$.
However, it would be unable to explain  
too hot and too cold objects (with $T_{\rm s}^\infty$ essentially
higher than the highest cooling curve
and essentially lower than the lowest curve
in Figs.\ \ref{fig9}-\ref{fig11}). 
A discovery of such hot or cold isolated
NSs would be of special interest.

\begin{acknowledgements}
One of the authors (DGY) gratefully acknowledges 
the support by the Heraeus foundation.
We are grateful to G.G.\ Pavlov for encouragement,
to M.E.\ Gusakov and
K.P.\ Levenfish, for useful comments.
The work was partially supported by RFBR (grants No.\ 02-02-17668
and 00-07-90183).
\end{acknowledgements}

\end{document}